\documentclass[twocolumn,showpacs,preprintnumbers,aps,prl, superscriptaddress]{revtex4-2}
\usepackage{hyperref}
\hypersetup{
    colorlinks =    true,
    linkcolor =     [rgb]{0.0,0.0,0.8},
    anchorcolor =   [rgb]{0.0,0.0,0.8},
    citecolor =     [rgb]{0.0,0.0,0.8},
    filecolor =     [rgb]{0.0,0.0,0.8},
    urlcolor =      [rgb]{0.0,0.0,0.8},
    pdftitle=       {Title},
    pdfsubject=     {Title},
    pdfauthor=      {A. Uthor}
}
\usepackage{nicefrac}
\usepackage{nccmath}
\usepackage[utf8]{inputenc}
\usepackage{amsmath}
\usepackage{graphicx}
\makeatletter
\usepackage{etoolbox} 
\usepackage{lipsum} 
\usepackage[capitalize]{cleveref}
\usepackage{nicefrac}
\usepackage{color, soul}
\usepackage{txfonts}
\usepackage{dcolumn}
\usepackage{bm}
\usepackage{natbib}
\usepackage{siunitx}
\usepackage{microtype}
\DeclareMathAlphabet\mathzapf       {T1}{pzc} {mb} {it}

\makeatother
\begin{document}

\title{A proposal for detecting the spin of a single electron in superfluid helium}

\author{Jinyong Ma}
\affiliation{Department of Physics, Yale University, New Haven, Connecticut 06520, USA}
\author{Y. S. S. Patil}
\affiliation{Department of Physics, Yale University, New Haven, Connecticut 06520, USA}
\author{Jiaxin Yu}
\affiliation{Department of Applied Physics, Yale University, New Haven, Connecticut 06520, USA}
\author{Yiqi Wang}
\affiliation{Department of Applied Physics, Yale University, New Haven, Connecticut 06520, USA}
\author{J. G. E. Harris$^{}$}\email{jack.harris@yale.edu}
\affiliation{Department of Physics, Yale University, New Haven, Connecticut 06520, USA}
\affiliation{Department of Applied Physics, Yale University, New Haven, Connecticut 06520, USA}
\affiliation{Yale Quantum Institute, Yale University, New Haven, Connecticut 06520, USA}

\date{\today}
\begin{abstract}
The electron bubble in superfluid helium has two degrees of freedom that may offer exceptionally low dissipation: the electron's spin and the bubble's motion. If these degrees of freedom can be read out and controlled with sufficient sensitivity, they would provide a novel platform for realizing a range of quantum technologies and for exploring open questions in the physics of superfluid helium. Here we propose a practical scheme for accomplishing this by trapping an electron bubble inside a superfluid-filled opto-acoustic cavity.
\end{abstract}


\maketitle

Atom-like defects in a solid can serve as quantum systems with low dissipation, strong coupling to external fields, and compatibility with well-established readout and control schemes. Defects that combine these features have been used in a range of applications, for example as quantum memories, single-photon sources, and as detectors that combine high sensitivity and spatial resolution. The performance of these systems is partially determined by the defect's internal properties, such as its level structure and its coupling to various fields. It is also determined by the properties of the host material, which represents a potential source of dissipation and noise \cite{Wolfowicz2021}. 

A number of defect systems have exhibited very long electron spin coherence times, including E$'$ centers in  SiO$_2$~\cite{RugarSingle2004}, N- or Si-vacancy centers in diamond~\cite{JelezkoObservation2004,schroderScalable2017}, and electrons trapped at Si/SiO$_2$ interfaces~\cite{XiaoElectrical2004}. These systems have been integrated with a range of readout schemes [including magnetic resonance force microscopy (MRFM)~\cite{RugarSingle2004}, nanoscale superconducting quantum interface devices~\cite{vasyukovScanning2013}, and laser excitation and fluorescence emission~\cite{schroderScalable2017}] and have facilitated the emergence of defect-based quantum sensing~\cite{taylorHighsensitivity2008}, computing~\cite{WeberQuantum2010} and information processing~\cite{WrachtrupProcessing2006, BradleyTenQubit2019}.

One atom-like defect that possesses a number of remarkable properties is the electron bubble (EB), which forms when a single electron is immersed in superfluid He. Such an electron resides inside a spherical bubble with radius \SI{1.9}{nm}~\cite{xingElectrons2020}. This system differs from the quantum defects found in solids in a number of respects. First, it is free to move, and at sufficiently low temperature it travels ballistically through the superfluid host. Second, the host material (i.e., superfluid He) is itself the topic of a number of outstanding questions that can be explored using EBs. These include the mechanical properties of vortices, and the onset and decay of turbulence~\cite{Popov1973,Duan1992,Duan1994,Baym1983,KozikKelvinWave2004, VinenKelvinWave2003,cheslerHolographic2013}.

Another remarkable feature of the EB is the exceptionally weak interaction between the electron's spin and the host material. Specifically, the absence of chemical and structural defects in superfluid $^4$He, together with the low concentration of nuclear spins (i.e., $^3$He impurities) are expected to provide an isolated EB with very long spin coherence. To date it has not been possible to measure the spin of an individual isolated EB, but measurements using large ensembles of transient (i.e., untrapped) EBs at $T>1$ K have found that the spin relaxation rate $T_1 \approx (100$ ms$)/ x_3$ (where $x_3$ is the fractional $^3$He concentration) for $0.5 \leq x_3 \leq 1$ \cite{ReichertMagneticresonance1983}. Naive extrapolation to natural He ($x_3 \approx 10^{-6}$) would imply $T_1 \approx 10^5$ s (consistent with measurements that place a lower bound $T_1 > 0.1$ s ~\cite{ZimmermannStudy1977}), and even longer for isotopically purified He.

Here we propose a scheme to efficiently detect and manipulate the spin of a single EB in superfluid He. In this scheme, the EB is trapped by an acoustic standing wave. A magnetic field gradient transduces the electron's spin state to the EB's motion within the trap, which in turn is monitored via an optical cavity. This approach is analogous to  MRFM (in the sample-on-cantilever configuration)~\cite{DegenNanoscale2009a}: here the sample is the electron, and the role of the cantilever is played by the trapped bubble. Many of the parameters of the device proposed here (such as its mechanical resonance frequency, quality factor, and operating temperature) are similar to those of cantilever-based MRFM devices~\cite{DegenNanoscale2009a}. However the EB's effective mass~\cite{huangEffective2017} is $\sim 10^{-14}$ that of typical MRFM cantilevers, which should result in a high signal-to-noise ratio (SNR) for measurements of the electron's spin. 

We emphasize that all of the necessary components for this scheme have been demonstrated \cite{McClintockField1969,KashkanovaSuperfluid2017a, MaminDetection2003,leirsetHeterodyne2013, ShkarinQuantum2019}, and that well-established models can be used to predict the device's performance. Details of these calculations are in Ref.~\cite{SI}.

\begin{figure}[htb] \centering
\includegraphics[width=0.45\textwidth]{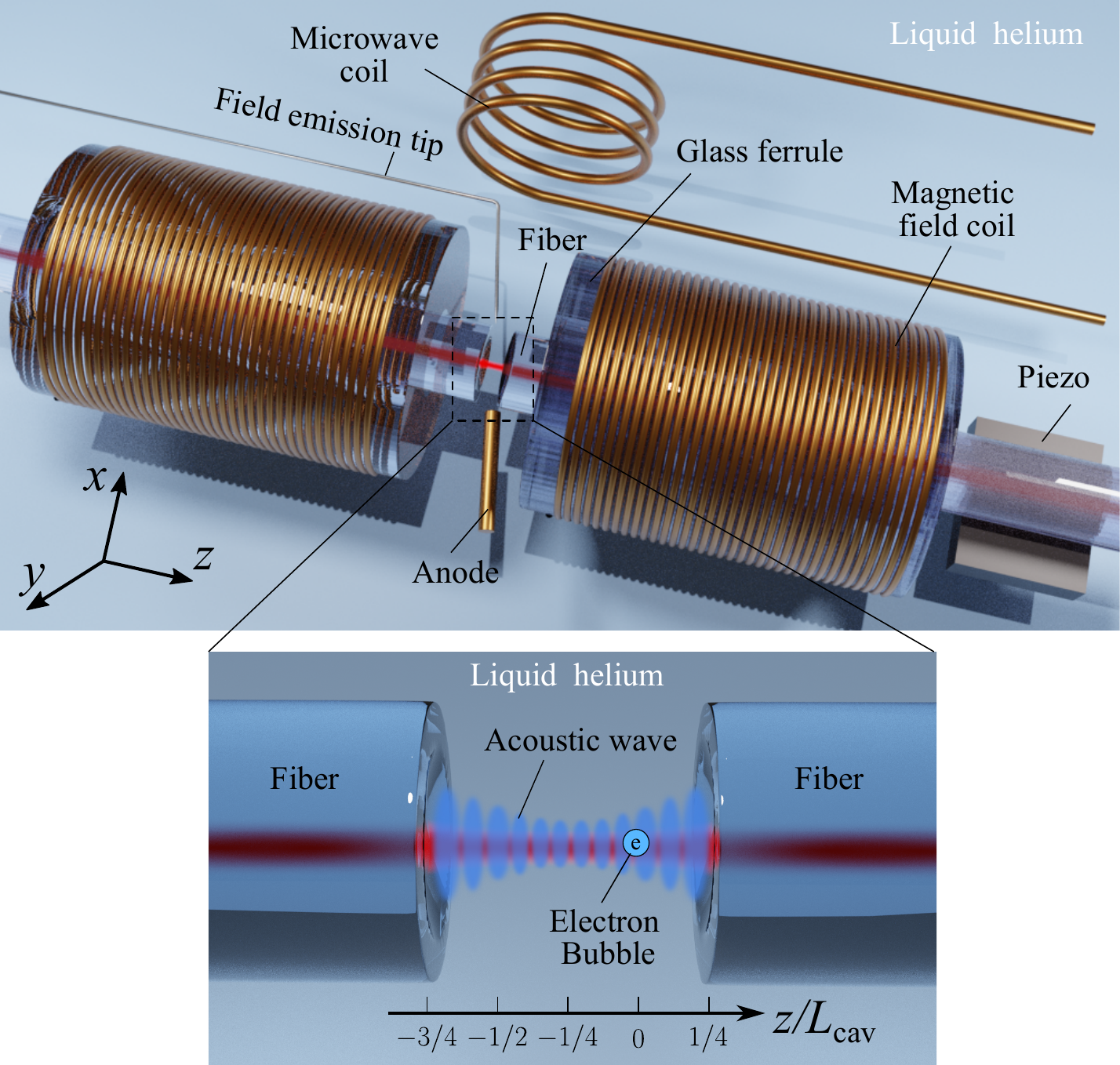} \caption{Schematic illustration of the proposed setup. The entire region is filled with superfluid He. An electron bubble (EB) is produced by field emission from a metal tip located near a cavity formed between two optical fibers. An acoustic standing wave in the cavity (inset, dark blue) traps the EB. A microwave coil is used to manipulate the electron's spin. A magnetic field gradient (generated by the field coil) transduces the state of electron's spin to the EB's displacement, which is measured using an optical mode of the cavity (inset, red). }
\label{fig:1}
\end{figure}

This scheme may be realized in a variety of ways, but for concreteness we focus on an implementation based on the superfluid-filled cavities described in Ref.~\cite{KashkanovaSuperfluid2017a, ShkarinQuantum2019} and illustrated in Fig.~\ref{fig:1}. The cavity (with length $L_{\mathrm{cav}} = 100$ \textmu m) is formed between the end faces of two optical fibers that have each been fabricated with a concave surface and coated with high-reflectivity dielectric mirrors~\cite{Hunger2012}. When filled with superfluid He at temperature $T<100$ mK, these cavities have been shown to confine optical modes with finesse $\mathcal{F} = 10^5$ (for vacuum wavelength $\lambda_{\mathrm{opt}} = 1550$ nm) and acoustic modes with quality factor $Q_{\mathrm{ac}} = 10^5$ (for wavelength $\lambda_{\mathrm{ac}} = 775$ nm). 

\begin{figure}[htb] \centering
\includegraphics[width=0.45\textwidth]{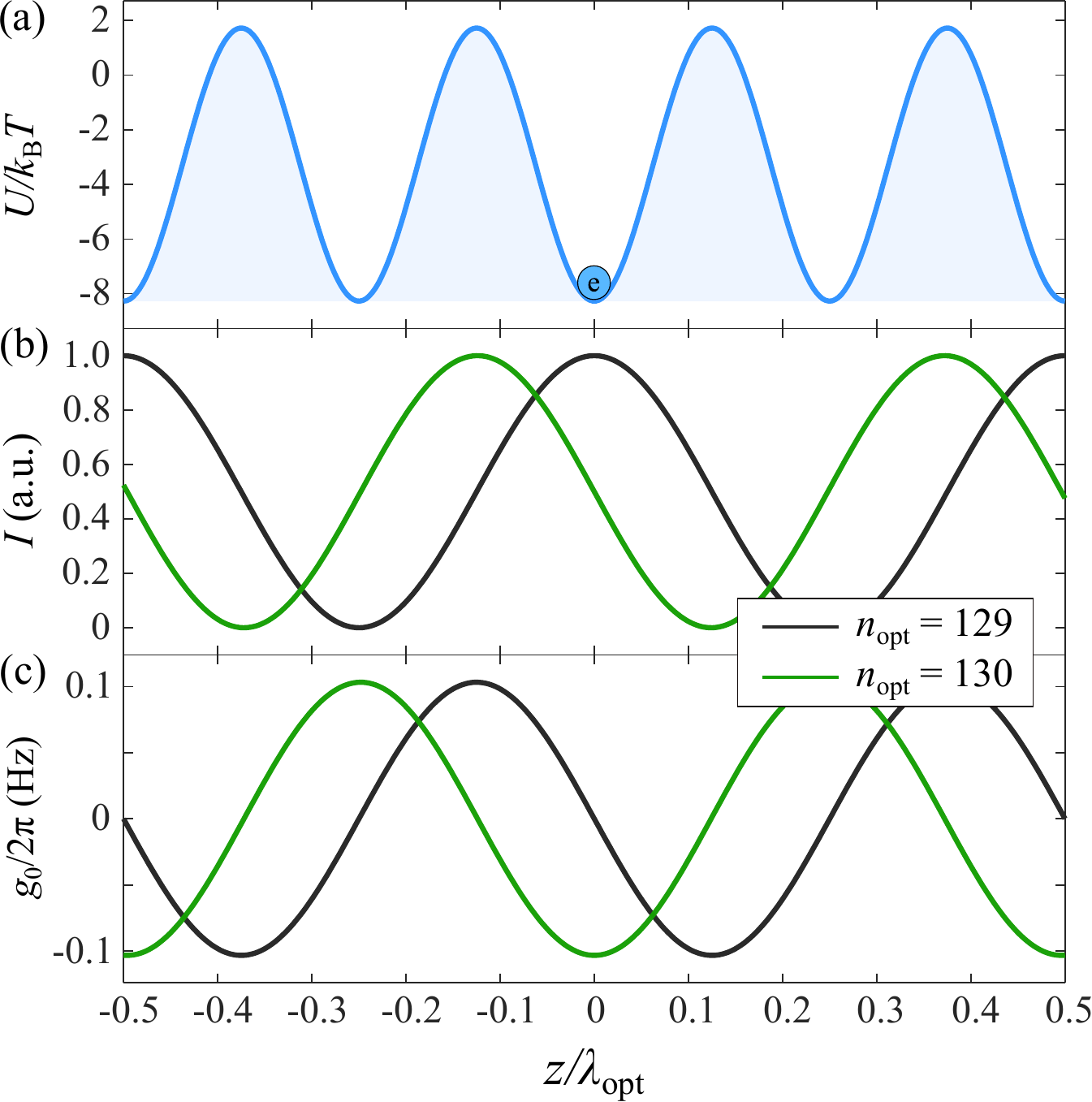} \caption{(a) The potential $U(z)$ of an EB in the acoustic standing wave. The blue circle indicates the EB's equilibrium position. (b) The intensity $I(z)$ of two optical modes. Black: the mode used to monitor the acoustic trap ($n_{\mathrm{opt}}=129$). Green: the mode used to detect the EB motion ($n_{\mathrm{opt}}=130$). (c) The optomechanical coupling rates $g_{0}$ of the two optical modes in (b). The horizontal axis denotes the EB position $z$ relative to the point one-quarter of the way between the cavity mirrors [see Fig. \ref{fig:1} (inset)].}
\label{fig:2}
\end{figure}

The EB is generated by field emission from a metal tip~\cite{McClintockField1969} positioned close to the cavity, and trapped by driving one of the cavity's acoustic modes to a large amplitude. As described in Ref.~\cite{SI}, the EB is attracted to the density antinodes of the resulting standing wave. Fig.~\ref{fig:2}(a) shows the potential energy $U(z)$ experienced by an EB in a standing wave with amplitude $\delta\rho_{\rm He}/\rho_{\rm He} = 2\times10^{-3}$ where $\delta\rho_{\rm He}$ is the change in the He density and $\rho_{\rm He} = 145$ kg/m$^3$ is the equilibrium density of superfluid He. This value of $\delta\rho_{\rm He}/\rho_{\rm He}$ is chosen to give a trap depth $U_{0} / k_{\mathrm{B}} = 300$ mK, a factor of $10$ greater than the operating temperature ($T = 30$ mK, as discussed below). It can be generated in an acoustic mode with $Q_{\mathrm{ac}} = 10^5$ and resonant frequency $\omega_{\mathrm{ac}}/2\pi =$ \SI{320}{MHz}~\cite{ShkarinQuantum2019} (corresponding to $\lambda_{\mathrm{ac}} = 775$ nm) by driving one of the fiber ends to an amplitude $\sim$ \SI{2}{fm} with an ultrasonic transducer~\cite{leirsetHeterodyne2013,Qifu2011,SI}. It results in a trap frequency (i.e., of the EB's center-of-mass motion)  $\omega_{\rm EB}/2\pi = 2.9$ MHz.  
 
The acoustic trap can be characterized even in the absence of an EB. This is because the density modulation produced by an acoustic mode detunes the cavity optical mode with $2n_{\rm opt}=n_{\rm ac}$, were $n_{\rm opt}$ ($n_{\rm ac}$) is the number of optical (acoustic) half-wavelengths fitting in the cavity (other optical modes only couple weakly to this acoustic mode~\cite{KashkanovaSuperfluid2017a}). For the particular device considered here, we take $n_\mathrm{ac} = 258$ and assume that the acoustic trap is monitored using the optical mode with $n_\mathrm{opt} = 129$. In this case the trap modulates the optical mode with depth $190\kappa$ (where $\kappa = 15$ MHz is the linewidth of the optical mode) which should be straightforward to measure~\cite{SI}.

As is common in levitated cavity optomechanics, the EB's displacement around the minimum of $U(z)$ is inferred from the detuning of  a ``probe'' optical mode \cite{Millen2020,SI}. This detuning occurs because the EB's index of refraction differs from that of the He filling the cavity, and so any optical mode is detuned by an amount proportional to its overlap with the EB~\cite{NimmrichterMaster2010, KieselCavity2013a}. The single-photon coupling rate~\cite{aspelmeyer_cavity-opto_review} $g_{0}$ that characterizes the strength of this interaction is optimized when the EB's equilibrium position coincides with the maximum gradient of the probe mode's intensity. For specificity, we consider an EB trapped a distance $L_{\rm cav}/4$ from one of the mirrors (see inset of Fig.~\ref{fig:1}), and monitored using the optical mode with $n_\mathrm{opt,probe} = 130$ as shown in Fig.~\ref{fig:2}(b,c). Since $n_\mathrm{opt,probe} \neq n_\mathrm{ac}/2$, this mode is first-order insensitive to the acoustic trap. 
 
\begin{figure}[htb] \centering
\includegraphics[width=0.45\textwidth]{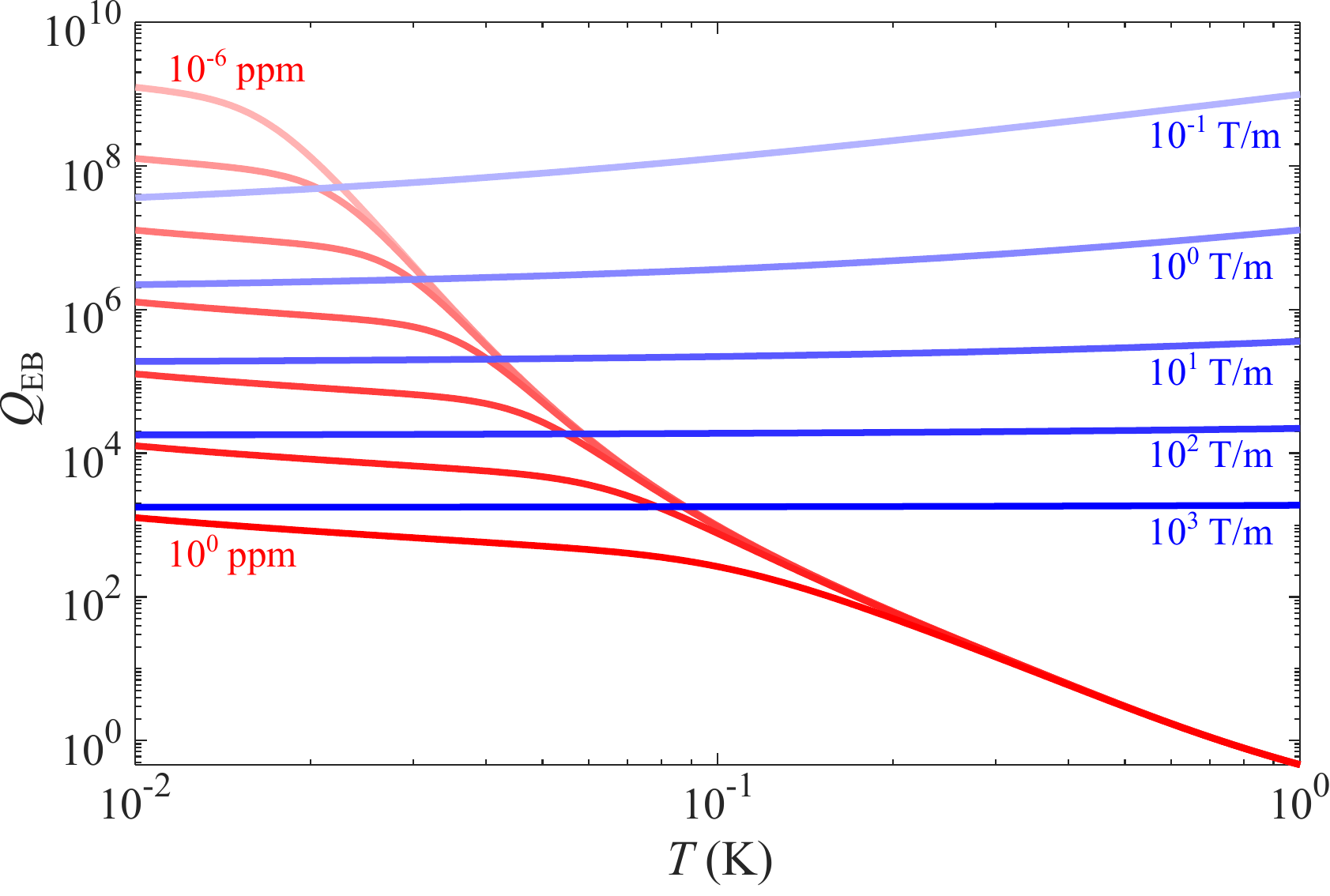} \caption{Requirements for sensing the spin of a single EB. Blue curves: the value of $Q_{\rm{EB}}$ required to obtain SNR $=$ 1 in a $1$ s measurement, assuming $P = 1$ \textmu W and $ \mathcal{F} = 10^5$. Each curve corresponds to the indicated value of $G$.  Red curves: $Q_{\rm{EB}}(T)$ for $x_3 = 10^{-6},10^{-5},...,10^{-1},1$ ppm. Blue and red curves cross at the value of $T$ below which the corresponding $G$ and $x_3$ would result in SNR $>1$. }
\label{fig:3}
\end{figure}

\begin{figure}[htb] 
\includegraphics[width=0.45\textwidth]{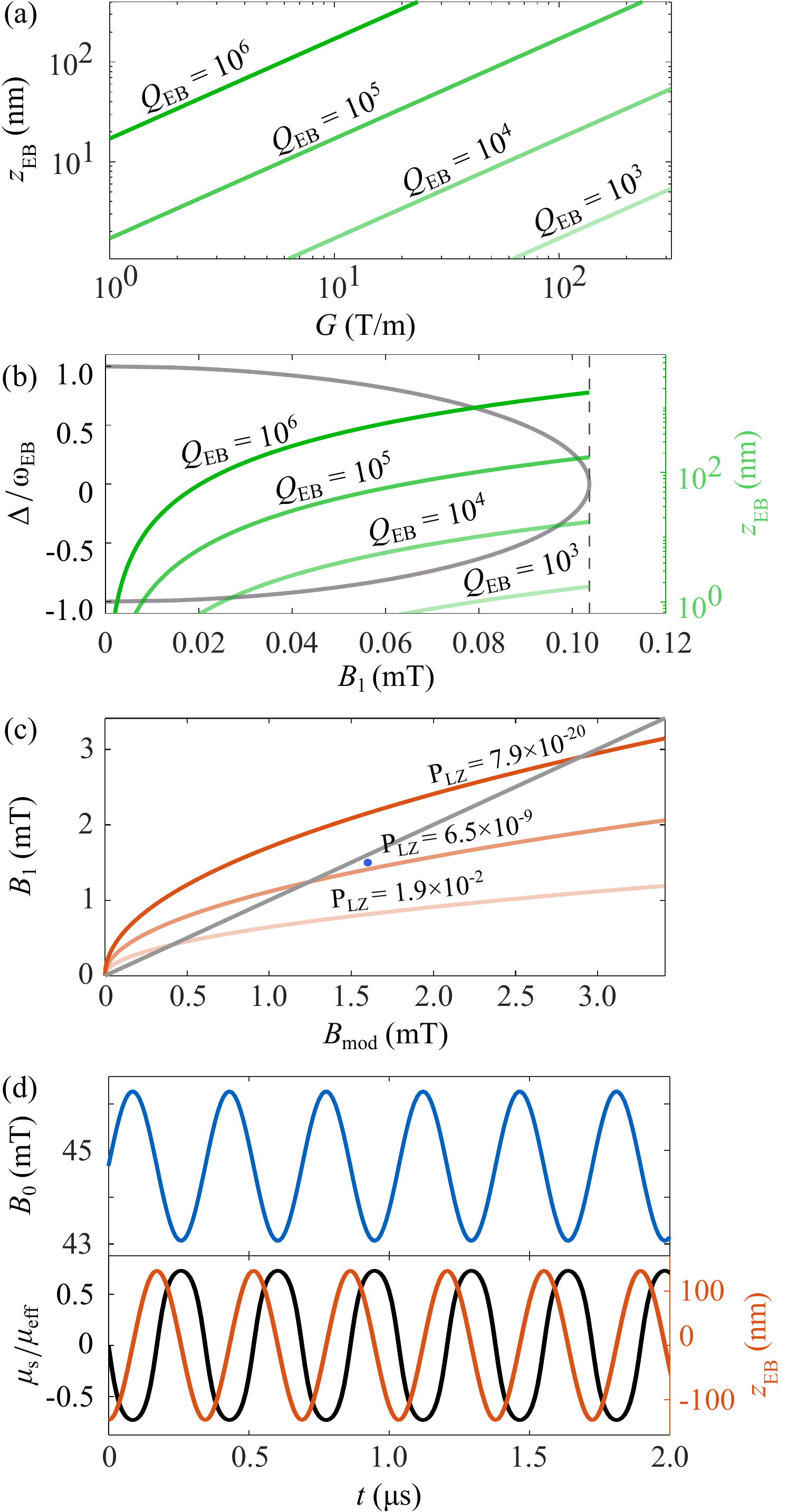} \caption{Three protocols for single spin detection. (a) Protocol I. The amplitude of the EB displacement $z_{\mathrm{EB}}(G)$. (b) Protocol II. Gray curve: the detuning $\Delta (B_1)$ that ensures the Rabi frequency is equal $\omega_{\mathrm{EB}}$ (for $B_0 = 45$ mT). Green curve: $z_{\mathrm{EB}}(B_1)$. Dashed line: the maximum value of $B_1$ for this protocol.  (c) Protocol III. Red curves: contours of constant error rates for the adiabatic passage. Gray line: the condition $B_1 = B_{\mathrm{mod}}$. (d) The timing for Protocol III for $B_1 = 1.5$ mT and $B_{\rm mod} = 1.6$ mT, indicated by the blue dot in (c). Panels (c),(d) assume $\bar{B}_0 = 45$ mT and $Q_{\rm EB} = 6 \times 10^5$.}
\label{fig:4}
\end{figure}

For the temperature range considered here ($T < 100$ mK), damping of the EB motion is primarily due to collisions with thermal phonons and impurity $^3$He atoms. Both processes are well-studied~\cite{BaymMobility1969,KramerLowTemperature1970, BorghesaniIons2007}, and the red curves in Fig.~\ref{fig:3} show the quality factor $Q_{\rm{EB}}(T)$ of the EB's motion calculated for an EB trapped in the potential of Fig.~\ref{fig:2}(a) \cite{SI}. At higher (lower) $T$ the damping is dominated by thermal phonons ($^3$He impurities). Fig.~\ref{fig:3} shows $Q_{\rm{EB}}(T)$ for $x_3$ ranging from $x_3 = 1$ ppm (comparable to the natural abundance) to the lowest level that is commercially available ($x_3 = 10^{-6}$ ppm).
 
The access to the EB's motion that is provided by the optical cavity can be used to study the spin of the EB. In particular, the state of the electron spin can be transduced by applying a magnetic field gradient of magnitude $G$, which results in a force of magnitude $F_{\rm mag} =  \mu_{\rm s}G$ on the EB, where $\mu_{\mathrm{s}}$ is the electron's magnetization along the gradient. In the scheme proposed here, a pair of coils (Fig.~\ref{fig:1}) produces a field along $z$ with magnitude $B_0$ and gradient (along $z$) $G$. A separate microwave coil is used to drive the electron's spin resonance. 

The spin is not assumed to be polarized in equilibrium; instead we assume (as in Ref.~\cite{RugarFluc}) that at any instant $\mu_{\rm s}= \pm g\mu_{\rm B}/2$ (where $\mu_{\rm B}$ is the Bohr magneton and $g=2.0023$ \cite{ReichertMagneticresonance1983}), and that the sign switches randomly with characteristic time $T_1$. In practice, this requires the averaging time of a measurement to be less than $T_1$.

We consider three protocols for measuring the EB spin. Each operates by causing $F_{\rm mag}$ to oscillate with frequency $\omega_{\mathrm{EB}}$ while measuring the EB's resulting motion. In Protocol I, $G$ is modulated while $\mu_{\rm s}$ is constant. In Protocols II \& III, $\mu_{\rm s}$ is modulated while $G$ is constant. In each case the amplitude of the signal (in the force domain) is simply $\mu_{\rm s}G$ (taking the static value of $\mu_{\rm s}$ and the oscillation amplitude of $G$ in Protocol I, and \textit{vice versa} in Protocols II \& III). In practice, each protocol is limited by different technical constraints, which we consider next. A more detailed description of each protocol is in Ref.~\cite{SI}. 

Protocol I is simply to modulate $G$ at frequency $\omega_{\rm EB}$. The amplitude of the resulting EB displacement $z_{\rm EB}$ is shown in Fig.~\ref{fig:4}(a). No microwave tone is applied. $B_0$ plays no direct role in this protocol. 

Protocol II uses the electron spin's Rabi flopping to produce oscillatory $F_{\rm mag}$. In this case $G$ and $B_0$ are fixed, and a microwave magnetic field with amplitude $B_1$ drives the transition between the spin states, whose Zeeman splitting is produced by $B_0$. This results in $\mu_{\rm s}(t)=\mu_{\rm B}\left(\frac{\omega_{1}}{\Omega}\right)^{2}\sin^{2}\left(\frac{\Omega t}{2}\right)$, where  $\Omega=(\Delta^{2}+\omega_{1}^{2})^{1/2}$ is the Rabi frequency, $\Delta$ is the detuning between the microwave drive and the Zeeman splitting, and $\omega_{1}=\frac{\mu_{\rm B}g}{\hbar}B_{1}$. Requiring $\Omega = \omega_{\rm EB}$ (to drive the EB motion resonantly) fixes $\Delta$ for a given $B_1$, as shown by the gray curve of Fig.~\ref{fig:4}(b). The resulting $z_{\rm EB}$ is shown by the green curves of Fig.~\ref{fig:4}(b). 

Protocol III is analogous to the one used in Refs.~\cite{MaminDetection2003,RugarSingle2004}. It uses constant $G$ and modulates the uniform field so that $B_0(t) = \bar{B}_0 + B_{\mathrm{mod}}\mathrm{sin}(\omega_{\mathrm{EB}}t)$. In addition, a fixed microwave tone with amplitude $B_1$ and frequency $\omega_0 =g \mu_{\mathrm{B}}\bar{B}_0/\hbar$ is applied. As a result, the Zeeman splitting passes through resonance with the microwaves twice per period of the modulation. If the condition for adiabaticity is met for each of these passages, the magnetization is given by~\cite{MaminDetection2003}

\begin{equation} \label{eq1}
    \mu_{s}(t)=\pm  \mu_{\rm eff} \frac{B_{{\rm mod}}\sin(\omega_{\mathrm{EB}}t)}{\left(B_{1}^{2}+B_{{\rm mod}}^{2}\sin^{2}(\omega_{\mathrm{EB}}t)\right)^{1/2}},
\end{equation}

\noindent where $\mu_{\rm eff} = g \mu_{\rm B}/2 $ and the overall sign flips randomly with characteristic time $T_1$. This signal is maximized when $B_{\mathrm{mod}} > B_1$ (the region to the right of the gray line in Fig.~\ref{fig:4}(c)). 

The probability that $\mu_{\mathrm{s}}(t)$ follows this form during an individual passage is $1 - P_\mathrm{LZ}$, where $P_\mathrm{LZ} = \mathrm{exp}( -\pi\mu_{\rm eff}B_{1}^{2} / \hbar\omega_{{\rm EB}}B_{\rm mod})$ is the well-known Landau-Zener formula \cite{SI}. The three red curves in Fig.~\ref{fig:4}(c) are contours of constant $P_{\mathrm{LZ}}(B_{1},B_{\rm mod})$. 
For $B_{1},B_{\rm mod}$ above the middle red curve, the rate of diabatic errors (i.e., in which $\mu_{\mathrm{s}}$ does not follow Eq.~\ref{eq1}) is less than $0.01$ s$^{-1}$. Fig.~\ref{fig:4}(d) shows the timing of this protocol for $B_1 = 1.5$ mT and $B_{\rm mod} = 1.6$ mT (corresponding to the blue point in Fig.~\ref{fig:4}(c)). 

\begin{figure}[htb] \centering
\includegraphics[width=0.45\textwidth]{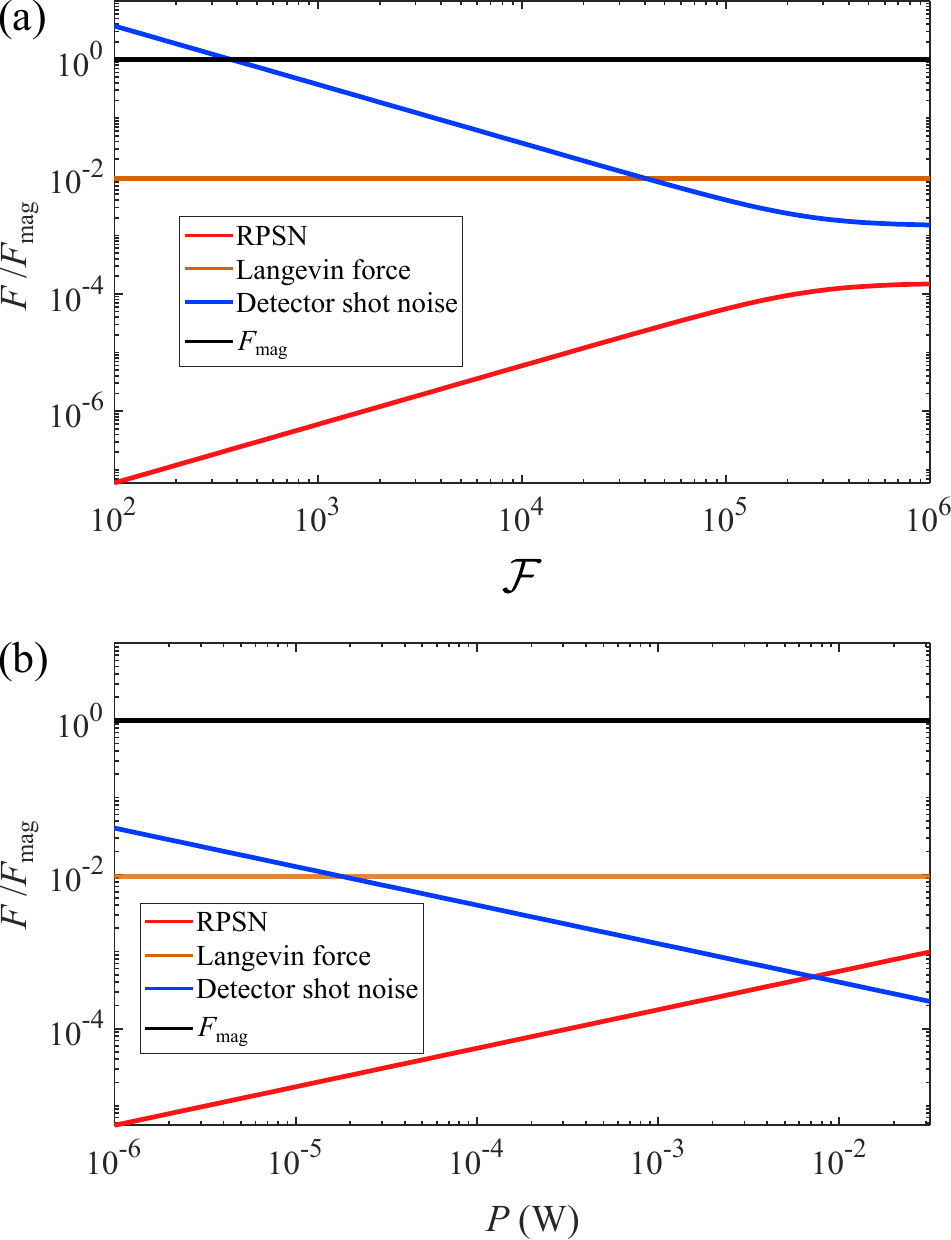} \caption{The estimated signal (black line) and noise (other lines) for single-spin detection. (a) Three noise sources (expressed as forces) compared with the signal $F_{\mathrm{mag}}$, all as a function of $\mathcal{F}$ and for $P = 100$ \textmu W. (b) The same as (a), but as a function of $P$ for $\mathcal{F} = 10^5$. Both (a) and (b) assume $x_3 = 10^{-3}$~ppm, $T = 30$~mK and $G = 100$~T/m.}
\label{fig:5}
\end{figure}

To estimate the feasibility of these protocols, we note that they all result in force signals of the same form (i.e., $F_{\rm mag}$), and are all subject to the same noise sources (see below). As a result, the same calculation of SNR applies to all three protocols. The dominant noise sources are expected to be: the Langevin force exerted by the thermal bath, the radiation pressure shot noise (RPSN) exerted by the photons in the optical mode, and the detector's shot noise (we assume the cavity detuning is monitored via heterodyne detection of the light from the cavity mode with $n_{\mathrm{opt}}=130$). These noise sources (as well as the heterodyne setup) are described in detail in Ref.~\cite{SI}. Typical contributions from each are shown in Fig.~\ref{fig:5}. 

For a wide range of experimental parameters, the dominant noise sources are the EB's thermal motion and the detector's shot noise. In this regime (where many MRFM experiments operate~\cite{DegenNanoscale2009a,RugarSingle2004,Grob2019}), the signal-to-noise ratio is $\mathrm{SNR} = F_{\mathrm{mag}}/ \sqrt{2}(F_{\rm th}+F_{\rm shot})$, where 

\begin{eqnarray}
F_{\rm th} &=& \sqrt{4k_{\mathrm{B}}Tm_{\rm EB}\omega_{\mathrm{EB}}b/Q_{\rm EB}} \\
F_{\rm shot} &=& \hbar \omega_{\rm EB}  \left|\kappa/2-i\omega_{\rm EB}\right| \frac{\kappa}{\kappa_{\rm{ex}}} \sqrt{ \frac{b m_{\rm{EB}} \omega_{\rm EB} \omega_{l}}{8 P Q_{\rm{EB}}^2 g_{0}^{2}} }.
\end{eqnarray}

\noindent Here $m_{\rm EB} = 1.6\times10^{-24}$ kg is the effective mass of the EB~\cite{huangEffective2017}, $b^{-1}$ is the averaging time, $P$ is the incident laser power, $\omega_l = 2 \pi c / \lambda_{\rm{opt}}$ and $\kappa_{\rm ex}$ is the external coupling rate of the optical cavity. All estimates presented here assume $\kappa_{\rm ex} = 0.44 \kappa$. They also assume $b^{-1}= 1$ s, which is much less than the expected $T_1$ (and, for Protocol III, much less than the predicted time between diabatic errors). 

The blue curves in Fig.~\ref{fig:3} show the value of $Q_{\rm EB}$ required to achieve SNR $=1$ (as a function of $T$ and $G$). They intersect the predicted $Q_{\rm EB}$ (red curves) at readily achievable values; for exmple, with $G=10^{2}$ T/m (three orders of magnitude less than in conventional MRFM experiments \cite{RugarSingle2004,DegenNanoscale2009a,Grob2019}) and $x_3 = 10^{-2}$ ppm, this intersection is at $T=55$ mK. For the same value of $G$, decreasing $x_3$ to $10^{-6}$ ppm and $T$ to $30$ mK would correspond to SNR $\approx 16$. 

In conclusion, we have proposed a scheme to manipulate and detect the spin of an individual EB trapped in an acousto-optical cavity filled with superfluid He. This approach would combine well-established techniques to produce, manipulate, and sense the electron spin, and should allow for the EB to be employed analogously to other atom-like defects.  The  unique features of the EB (in particular, its mobility and long spin coherence) may open new opportunities in high-sensitivity magnetometry, and in the study of superfluid He. 

This scheme may be extended to create a linear array of EBs by trapping each one in a different minimum of $U(z)$. In the presence of a magnetic field gradient, the EBs' Coulomb interaction would be spin-dependent, and so could provide access to phenomena that are typically studied in trapped ion arrays \cite{Molmer1999}. 

Additionally, it may be possible to apply the cavity-enhanced acoustic trapping and optical readout proposed here to other impurities whose size and acoustic contrast are not too different from the EB. One example would be bubbles containing multiple electrons \cite{Volodin1977,Yadav2021}. Another would be the wide range of molecules, clusters, and nanoparticles that are commonly immersed in superfluid He for high-precision spectroscopic measurements \cite{slenczka2022molecules,Toennies2004}. Stably trapping these impurities may allow for more detailed studies than have been possible to date.

\vspace{4mm}

This work is supported by the AFOSR (Grant No. FA9550-21-1-0152), the Vannevar Bush Faculty Fellowship (No. N00014-20-1-2628), the Quantum Information Science Enabled Discovery (QuantISED) for High Energy Physics (KA2401032), ONR MURI on Quantum Optomechanics (Grant No. N00014-15-1-2761), and the NSF (Grant No. 1707703). This material is based upon work supported by the Air Force Office of Scientific Research and the Office of Naval Research under award number FA9550-23-1-0333.

\bibliographystyle{apsrev4-2}
\bibliography{ms}

\begin{thebibliography}{44}%
\makeatletter
\providecommand \@ifxundefined [1]{%
 \@ifx{#1\undefined}
}%
\providecommand \@ifnum [1]{%
 \ifnum #1\expandafter \@firstoftwo
 \else \expandafter \@secondoftwo
 \fi
}%
\providecommand \@ifx [1]{%
 \ifx #1\expandafter \@firstoftwo
 \else \expandafter \@secondoftwo
 \fi
}%
\providecommand \natexlab [1]{#1}%
\providecommand \enquote  [1]{``#1''}%
\providecommand \bibnamefont  [1]{#1}%
\providecommand \bibfnamefont [1]{#1}%
\providecommand \citenamefont [1]{#1}%
\providecommand \href@noop [0]{\@secondoftwo}%
\providecommand \href [0]{\begingroup \@sanitize@url \@href}%
\providecommand \@href[1]{\@@startlink{#1}\@@href}%
\providecommand \@@href[1]{\endgroup#1\@@endlink}%
\providecommand \@sanitize@url [0]{\catcode `\\12\catcode `\$12\catcode
  `\&12\catcode `\#12\catcode `\^12\catcode `\_12\catcode `\%12\relax}%
\providecommand \@@startlink[1]{}%
\providecommand \@@endlink[0]{}%
\providecommand \url  [0]{\begingroup\@sanitize@url \@url }%
\providecommand \@url [1]{\endgroup\@href {#1}{\urlprefix }}%
\providecommand \urlprefix  [0]{URL }%
\providecommand \Eprint [0]{\href }%
\providecommand \doibase [0]{https://doi.org/}%
\providecommand \selectlanguage [0]{\@gobble}%
\providecommand \bibinfo  [0]{\@secondoftwo}%
\providecommand \bibfield  [0]{\@secondoftwo}%
\providecommand \translation [1]{[#1]}%
\providecommand \BibitemOpen [0]{}%
\providecommand \bibitemStop [0]{}%
\providecommand \bibitemNoStop [0]{.\EOS\space}%
\providecommand \EOS [0]{\spacefactor3000\relax}%
\providecommand \BibitemShut  [1]{\csname bibitem#1\endcsname}%
\let\auto@bib@innerbib\@empty
\bibitem [{\citenamefont {Wolfowicz}\ \emph {et~al.}(2021)\citenamefont
  {Wolfowicz}, \citenamefont {Heremans}, \citenamefont {Anderson},
  \citenamefont {Kanai}, \citenamefont {Seo}, \citenamefont {Gali},
  \citenamefont {Galli},\ and\ \citenamefont {Awschalom}}]{Wolfowicz2021}%
  \BibitemOpen
  \bibfield  {author} {\bibinfo {author} {\bibfnamefont {G.}~\bibnamefont
  {Wolfowicz}}, \bibinfo {author} {\bibfnamefont {F.~J.}\ \bibnamefont
  {Heremans}}, \bibinfo {author} {\bibfnamefont {C.~P.}\ \bibnamefont
  {Anderson}}, \bibinfo {author} {\bibfnamefont {S.}~\bibnamefont {Kanai}},
  \bibinfo {author} {\bibfnamefont {H.}~\bibnamefont {Seo}}, \bibinfo {author}
  {\bibfnamefont {A.}~\bibnamefont {Gali}}, \bibinfo {author} {\bibfnamefont
  {G.}~\bibnamefont {Galli}},\ and\ \bibinfo {author} {\bibfnamefont {D.~D.}\
  \bibnamefont {Awschalom}},\ }\href
  {https://doi.org/10.1038/s41578-021-00306-y} {\bibfield  {journal} {\bibinfo
  {journal} {Nature Reviews Materials}\ }\textbf {\bibinfo {volume} {6}},\
  \bibinfo {pages} {906} (\bibinfo {year} {2021})}\BibitemShut {NoStop}%
\bibitem [{\citenamefont {Rugar}\ \emph {et~al.}(2004)\citenamefont {Rugar},
  \citenamefont {Budakian}, \citenamefont {Mamin},\ and\ \citenamefont
  {Chui}}]{RugarSingle2004}%
  \BibitemOpen
  \bibfield  {author} {\bibinfo {author} {\bibfnamefont {D.}~\bibnamefont
  {Rugar}}, \bibinfo {author} {\bibfnamefont {R.}~\bibnamefont {Budakian}},
  \bibinfo {author} {\bibfnamefont {H.~J.}\ \bibnamefont {Mamin}},\ and\
  \bibinfo {author} {\bibfnamefont {B.~W.}\ \bibnamefont {Chui}},\ }\href
  {https://doi.org/10.1038/nature02658} {\bibfield  {journal} {\bibinfo
  {journal} {Nature}\ }\textbf {\bibinfo {volume} {430}},\ \bibinfo {pages}
  {329} (\bibinfo {year} {2004})}\BibitemShut {NoStop}%
\bibitem [{\citenamefont {Jelezko}\ \emph {et~al.}(2004)\citenamefont
  {Jelezko}, \citenamefont {Gaebel}, \citenamefont {Popa}, \citenamefont
  {Gruber},\ and\ \citenamefont {Wrachtrup}}]{JelezkoObservation2004}%
  \BibitemOpen
  \bibfield  {author} {\bibinfo {author} {\bibfnamefont {F.}~\bibnamefont
  {Jelezko}}, \bibinfo {author} {\bibfnamefont {T.}~\bibnamefont {Gaebel}},
  \bibinfo {author} {\bibfnamefont {I.}~\bibnamefont {Popa}}, \bibinfo {author}
  {\bibfnamefont {A.}~\bibnamefont {Gruber}},\ and\ \bibinfo {author}
  {\bibfnamefont {J.}~\bibnamefont {Wrachtrup}},\ }\href
  {https://doi.org/10.1103/PhysRevLett.92.076401} {\bibfield  {journal}
  {\bibinfo  {journal} {Physical Review Letters}\ }\textbf {\bibinfo {volume}
  {92}},\ \bibinfo {pages} {076401} (\bibinfo {year} {2004})}\BibitemShut
  {NoStop}%
\bibitem [{\citenamefont {Schr{\"o}der}\ \emph {et~al.}(2017)\citenamefont
  {Schr{\"o}der}, \citenamefont {Trusheim}, \citenamefont {Walsh},
  \citenamefont {Li}, \citenamefont {Zheng}, \citenamefont {Schukraft},
  \citenamefont {Sipahigil}, \citenamefont {Evans}, \citenamefont {Sukachev},
  \citenamefont {Nguyen}, \citenamefont {Pacheco}, \citenamefont {Camacho},
  \citenamefont {Bielejec}, \citenamefont {Lukin},\ and\ \citenamefont
  {Englund}}]{schroderScalable2017}%
  \BibitemOpen
  \bibfield  {author} {\bibinfo {author} {\bibfnamefont {T.}~\bibnamefont
  {Schr{\"o}der}}, \bibinfo {author} {\bibfnamefont {M.~E.}\ \bibnamefont
  {Trusheim}}, \bibinfo {author} {\bibfnamefont {M.}~\bibnamefont {Walsh}},
  \bibinfo {author} {\bibfnamefont {L.}~\bibnamefont {Li}}, \bibinfo {author}
  {\bibfnamefont {J.}~\bibnamefont {Zheng}}, \bibinfo {author} {\bibfnamefont
  {M.}~\bibnamefont {Schukraft}}, \bibinfo {author} {\bibfnamefont
  {A.}~\bibnamefont {Sipahigil}}, \bibinfo {author} {\bibfnamefont {R.~E.}\
  \bibnamefont {Evans}}, \bibinfo {author} {\bibfnamefont {D.~D.}\ \bibnamefont
  {Sukachev}}, \bibinfo {author} {\bibfnamefont {C.~T.}\ \bibnamefont
  {Nguyen}}, \bibinfo {author} {\bibfnamefont {J.~L.}\ \bibnamefont {Pacheco}},
  \bibinfo {author} {\bibfnamefont {R.~M.}\ \bibnamefont {Camacho}}, \bibinfo
  {author} {\bibfnamefont {E.~S.}\ \bibnamefont {Bielejec}}, \bibinfo {author}
  {\bibfnamefont {M.~D.}\ \bibnamefont {Lukin}},\ and\ \bibinfo {author}
  {\bibfnamefont {D.}~\bibnamefont {Englund}},\ }\href
  {https://doi.org/10.1038/ncomms15376} {\bibfield  {journal} {\bibinfo
  {journal} {Nature Communications}\ }\textbf {\bibinfo {volume} {8}},\
  \bibinfo {pages} {15376} (\bibinfo {year} {2017})}\BibitemShut {NoStop}%
\bibitem [{\citenamefont {Xiao}\ \emph {et~al.}(2004)\citenamefont {Xiao},
  \citenamefont {Martin}, \citenamefont {Yablonovitch},\ and\ \citenamefont
  {Jiang}}]{XiaoElectrical2004}%
  \BibitemOpen
  \bibfield  {author} {\bibinfo {author} {\bibfnamefont {M.}~\bibnamefont
  {Xiao}}, \bibinfo {author} {\bibfnamefont {I.}~\bibnamefont {Martin}},
  \bibinfo {author} {\bibfnamefont {E.}~\bibnamefont {Yablonovitch}},\ and\
  \bibinfo {author} {\bibfnamefont {H.~W.}\ \bibnamefont {Jiang}},\ }\href
  {https://doi.org/10.1038/nature02727} {\bibfield  {journal} {\bibinfo
  {journal} {Nature}\ }\textbf {\bibinfo {volume} {430}},\ \bibinfo {pages}
  {435} (\bibinfo {year} {2004})}\BibitemShut {NoStop}%
\bibitem [{\citenamefont {Vasyukov}\ \emph {et~al.}(2013)\citenamefont
  {Vasyukov}, \citenamefont {Anahory}, \citenamefont {Embon}, \citenamefont
  {Halbertal}, \citenamefont {Cuppens}, \citenamefont {Neeman}, \citenamefont
  {Finkler}, \citenamefont {Segev}, \citenamefont {Myasoedov}, \citenamefont
  {Rappaport}, \citenamefont {Huber},\ and\ \citenamefont
  {Zeldov}}]{vasyukovScanning2013}%
  \BibitemOpen
  \bibfield  {author} {\bibinfo {author} {\bibfnamefont {D.}~\bibnamefont
  {Vasyukov}}, \bibinfo {author} {\bibfnamefont {Y.}~\bibnamefont {Anahory}},
  \bibinfo {author} {\bibfnamefont {L.}~\bibnamefont {Embon}}, \bibinfo
  {author} {\bibfnamefont {D.}~\bibnamefont {Halbertal}}, \bibinfo {author}
  {\bibfnamefont {J.}~\bibnamefont {Cuppens}}, \bibinfo {author} {\bibfnamefont
  {L.}~\bibnamefont {Neeman}}, \bibinfo {author} {\bibfnamefont
  {A.}~\bibnamefont {Finkler}}, \bibinfo {author} {\bibfnamefont
  {Y.}~\bibnamefont {Segev}}, \bibinfo {author} {\bibfnamefont
  {Y.}~\bibnamefont {Myasoedov}}, \bibinfo {author} {\bibfnamefont {M.~L.}\
  \bibnamefont {Rappaport}}, \bibinfo {author} {\bibfnamefont {M.~E.}\
  \bibnamefont {Huber}},\ and\ \bibinfo {author} {\bibfnamefont
  {E.}~\bibnamefont {Zeldov}},\ }\href {https://doi.org/10.1038/nnano.2013.169}
  {\bibfield  {journal} {\bibinfo  {journal} {Nature Nanotechnology}\ }\textbf
  {\bibinfo {volume} {8}},\ \bibinfo {pages} {639} (\bibinfo {year}
  {2013})}\BibitemShut {NoStop}%
\bibitem [{\citenamefont {Taylor}\ \emph {et~al.}(2008)\citenamefont {Taylor},
  \citenamefont {Cappellaro}, \citenamefont {Childress}, \citenamefont {Jiang},
  \citenamefont {Budker}, \citenamefont {Hemmer}, \citenamefont {Yacoby},
  \citenamefont {Walsworth},\ and\ \citenamefont
  {Lukin}}]{taylorHighsensitivity2008}%
  \BibitemOpen
  \bibfield  {author} {\bibinfo {author} {\bibfnamefont {J.~M.}\ \bibnamefont
  {Taylor}}, \bibinfo {author} {\bibfnamefont {P.}~\bibnamefont {Cappellaro}},
  \bibinfo {author} {\bibfnamefont {L.}~\bibnamefont {Childress}}, \bibinfo
  {author} {\bibfnamefont {L.}~\bibnamefont {Jiang}}, \bibinfo {author}
  {\bibfnamefont {D.}~\bibnamefont {Budker}}, \bibinfo {author} {\bibfnamefont
  {P.~R.}\ \bibnamefont {Hemmer}}, \bibinfo {author} {\bibfnamefont
  {A.}~\bibnamefont {Yacoby}}, \bibinfo {author} {\bibfnamefont
  {R.}~\bibnamefont {Walsworth}},\ and\ \bibinfo {author} {\bibfnamefont
  {M.~D.}\ \bibnamefont {Lukin}},\ }\href {https://doi.org/10.1038/nphys1075}
  {\bibfield  {journal} {\bibinfo  {journal} {Nature Physics}\ }\textbf
  {\bibinfo {volume} {4}},\ \bibinfo {pages} {810} (\bibinfo {year}
  {2008})}\BibitemShut {NoStop}%
\bibitem [{\citenamefont {Weber}\ \emph {et~al.}(2010)\citenamefont {Weber},
  \citenamefont {Koehl}, \citenamefont {Varley}, \citenamefont {Janotti},
  \citenamefont {Buckley}, \citenamefont {{Van de Walle}},\ and\ \citenamefont
  {Awschalom}}]{WeberQuantum2010}%
  \BibitemOpen
  \bibfield  {author} {\bibinfo {author} {\bibfnamefont {J.~R.}\ \bibnamefont
  {Weber}}, \bibinfo {author} {\bibfnamefont {W.~F.}\ \bibnamefont {Koehl}},
  \bibinfo {author} {\bibfnamefont {J.~B.}\ \bibnamefont {Varley}}, \bibinfo
  {author} {\bibfnamefont {A.}~\bibnamefont {Janotti}}, \bibinfo {author}
  {\bibfnamefont {B.~B.}\ \bibnamefont {Buckley}}, \bibinfo {author}
  {\bibfnamefont {C.~G.}\ \bibnamefont {{Van de Walle}}},\ and\ \bibinfo
  {author} {\bibfnamefont {D.~D.}\ \bibnamefont {Awschalom}},\ }\href
  {https://doi.org/10.1073/pnas.1003052107} {\bibfield  {journal} {\bibinfo
  {journal} {Proceedings of the National Academy of Sciences}\ }\textbf
  {\bibinfo {volume} {107}},\ \bibinfo {pages} {8513} (\bibinfo {year}
  {2010})}\BibitemShut {NoStop}%
\bibitem [{\citenamefont {Wrachtrup}\ and\ \citenamefont
  {Jelezko}(2006)}]{WrachtrupProcessing2006}%
  \BibitemOpen
  \bibfield  {author} {\bibinfo {author} {\bibfnamefont {J.}~\bibnamefont
  {Wrachtrup}}\ and\ \bibinfo {author} {\bibfnamefont {F.}~\bibnamefont
  {Jelezko}},\ }\href {https://doi.org/10.1088/0953-8984/18/21/S08} {\bibfield
  {journal} {\bibinfo  {journal} {Journal of Physics: Condensed Matter}\
  }\textbf {\bibinfo {volume} {18}},\ \bibinfo {pages} {S807} (\bibinfo {year}
  {2006})}\BibitemShut {NoStop}%
\bibitem [{\citenamefont {Bradley}\ \emph {et~al.}(2019)\citenamefont
  {Bradley}, \citenamefont {Randall}, \citenamefont {Abobeih}, \citenamefont
  {Berrevoets}, \citenamefont {Degen}, \citenamefont {Bakker}, \citenamefont
  {Markham}, \citenamefont {Twitchen},\ and\ \citenamefont
  {Taminiau}}]{BradleyTenQubit2019}%
  \BibitemOpen
  \bibfield  {author} {\bibinfo {author} {\bibfnamefont {C.~E.}\ \bibnamefont
  {Bradley}}, \bibinfo {author} {\bibfnamefont {J.}~\bibnamefont {Randall}},
  \bibinfo {author} {\bibfnamefont {M.~H.}\ \bibnamefont {Abobeih}}, \bibinfo
  {author} {\bibfnamefont {R.~C.}\ \bibnamefont {Berrevoets}}, \bibinfo
  {author} {\bibfnamefont {M.~J.}\ \bibnamefont {Degen}}, \bibinfo {author}
  {\bibfnamefont {M.~A.}\ \bibnamefont {Bakker}}, \bibinfo {author}
  {\bibfnamefont {M.}~\bibnamefont {Markham}}, \bibinfo {author} {\bibfnamefont
  {D.~J.}\ \bibnamefont {Twitchen}},\ and\ \bibinfo {author} {\bibfnamefont
  {T.~H.}\ \bibnamefont {Taminiau}},\ }\href
  {https://doi.org/10.1103/PhysRevX.9.031045} {\bibfield  {journal} {\bibinfo
  {journal} {Physical Review X}\ }\textbf {\bibinfo {volume} {9}},\ \bibinfo
  {pages} {031045} (\bibinfo {year} {2019})}\BibitemShut {NoStop}%
\bibitem [{\citenamefont {Xing}\ and\ \citenamefont
  {Maris}(2020)}]{xingElectrons2020}%
  \BibitemOpen
  \bibfield  {author} {\bibinfo {author} {\bibfnamefont {Y.}~\bibnamefont
  {Xing}}\ and\ \bibinfo {author} {\bibfnamefont {H.~J.}\ \bibnamefont
  {Maris}},\ }\href {https://doi.org/10.1007/s10909-020-02422-5} {\bibfield
  {journal} {\bibinfo  {journal} {Journal of Low Temperature Physics}\ }\textbf
  {\bibinfo {volume} {201}},\ \bibinfo {pages} {634} (\bibinfo {year}
  {2020})}\BibitemShut {NoStop}%
\bibitem [{\citenamefont {Popov}(1973)}]{Popov1973}%
  \BibitemOpen
  \bibfield  {author} {\bibinfo {author} {\bibfnamefont {V.~N.}\ \bibnamefont
  {Popov}},\ }\href {http://jetp.ras.ru/cgi-bin/e/index/e/37/2/p341?a=list}
  {\bibfield  {journal} {\bibinfo  {journal} {Sov. Phys. JETP}\ }\textbf
  {\bibinfo {volume} {37}} (\bibinfo {year} {1973})}\BibitemShut {NoStop}%
\bibitem [{\citenamefont {Duan}\ and\ \citenamefont
  {Leggett}(1992)}]{Duan1992}%
  \BibitemOpen
  \bibfield  {author} {\bibinfo {author} {\bibfnamefont {J.-M.}\ \bibnamefont
  {Duan}}\ and\ \bibinfo {author} {\bibfnamefont {A.~J.}\ \bibnamefont
  {Leggett}},\ }\href {https://doi.org/10.1103/PhysRevLett.68.1216} {\bibfield
  {journal} {\bibinfo  {journal} {Phys. Rev. Lett.}\ }\textbf {\bibinfo
  {volume} {68}},\ \bibinfo {pages} {1216} (\bibinfo {year}
  {1992})}\BibitemShut {NoStop}%
\bibitem [{\citenamefont {Duan}(1994)}]{Duan1994}%
  \BibitemOpen
  \bibfield  {author} {\bibinfo {author} {\bibfnamefont {J.-M.}\ \bibnamefont
  {Duan}},\ }\href {https://doi.org/10.1103/PhysRevB.49.12381} {\bibfield
  {journal} {\bibinfo  {journal} {Phys. Rev. B}\ }\textbf {\bibinfo {volume}
  {49}},\ \bibinfo {pages} {12381} (\bibinfo {year} {1994})}\BibitemShut
  {NoStop}%
\bibitem [{\citenamefont {Baym}\ and\ \citenamefont
  {Chandler}(1983)}]{Baym1983}%
  \BibitemOpen
  \bibfield  {author} {\bibinfo {author} {\bibfnamefont {G.}~\bibnamefont
  {Baym}}\ and\ \bibinfo {author} {\bibfnamefont {E.}~\bibnamefont
  {Chandler}},\ }\href {https://doi.org/10.1007/BF00681839} {\bibfield
  {journal} {\bibinfo  {journal} {Journal of Low Temperature Physics}\ }\textbf
  {\bibinfo {volume} {50}},\ \bibinfo {pages} {57} (\bibinfo {year}
  {1983})}\BibitemShut {NoStop}%
\bibitem [{\citenamefont {Kozik}\ and\ \citenamefont
  {Svistunov}(2004)}]{KozikKelvinWave2004}%
  \BibitemOpen
  \bibfield  {author} {\bibinfo {author} {\bibfnamefont {E.}~\bibnamefont
  {Kozik}}\ and\ \bibinfo {author} {\bibfnamefont {B.}~\bibnamefont
  {Svistunov}},\ }\href {https://doi.org/10.1103/PhysRevLett.92.035301}
  {\bibfield  {journal} {\bibinfo  {journal} {Physical Review Letters}\
  }\textbf {\bibinfo {volume} {92}},\ \bibinfo {pages} {035301} (\bibinfo
  {year} {2004})}\BibitemShut {NoStop}%
\bibitem [{\citenamefont {Vinen}\ \emph {et~al.}(2003)\citenamefont {Vinen},
  \citenamefont {Tsubota},\ and\ \citenamefont {Mitani}}]{VinenKelvinWave2003}%
  \BibitemOpen
  \bibfield  {author} {\bibinfo {author} {\bibfnamefont {W.~F.}\ \bibnamefont
  {Vinen}}, \bibinfo {author} {\bibfnamefont {M.}~\bibnamefont {Tsubota}},\
  and\ \bibinfo {author} {\bibfnamefont {A.}~\bibnamefont {Mitani}},\ }\href
  {https://doi.org/10.1103/PhysRevLett.91.135301} {\bibfield  {journal}
  {\bibinfo  {journal} {Physical Review Letters}\ }\textbf {\bibinfo {volume}
  {91}},\ \bibinfo {pages} {135301} (\bibinfo {year} {2003})}\BibitemShut
  {NoStop}%
\bibitem [{\citenamefont {Chesler}\ \emph {et~al.}(2013)\citenamefont
  {Chesler}, \citenamefont {Liu},\ and\ \citenamefont
  {Adams}}]{cheslerHolographic2013}%
  \BibitemOpen
  \bibfield  {author} {\bibinfo {author} {\bibfnamefont {P.~M.}\ \bibnamefont
  {Chesler}}, \bibinfo {author} {\bibfnamefont {H.}~\bibnamefont {Liu}},\ and\
  \bibinfo {author} {\bibfnamefont {A.}~\bibnamefont {Adams}},\ }\href
  {https://doi.org/10.1126/science.1233529} {\bibfield  {journal} {\bibinfo
  {journal} {Science}\ }\textbf {\bibinfo {volume} {341}},\ \bibinfo {pages}
  {368} (\bibinfo {year} {2013})}\BibitemShut {NoStop}%
\bibitem [{\citenamefont {Reichert}\ and\ \citenamefont
  {Jarosik}(1983)}]{ReichertMagneticresonance1983}%
  \BibitemOpen
  \bibfield  {author} {\bibinfo {author} {\bibfnamefont {J.~F.}\ \bibnamefont
  {Reichert}}\ and\ \bibinfo {author} {\bibfnamefont {N.~C.}\ \bibnamefont
  {Jarosik}},\ }\href {https://doi.org/10.1103/PhysRevB.27.2710} {\bibfield
  {journal} {\bibinfo  {journal} {Physical Review B}\ }\textbf {\bibinfo
  {volume} {27}},\ \bibinfo {pages} {2710} (\bibinfo {year}
  {1983})}\BibitemShut {NoStop}%
\bibitem [{\citenamefont {Zimmermann}\ \emph {et~al.}(1977)\citenamefont
  {Zimmermann}, \citenamefont {Reichert},\ and\ \citenamefont
  {Dahm}}]{ZimmermannStudy1977}%
  \BibitemOpen
  \bibfield  {author} {\bibinfo {author} {\bibfnamefont {P.~H.}\ \bibnamefont
  {Zimmermann}}, \bibinfo {author} {\bibfnamefont {J.~F.}\ \bibnamefont
  {Reichert}},\ and\ \bibinfo {author} {\bibfnamefont {A.~J.}\ \bibnamefont
  {Dahm}},\ }\href {https://doi.org/10.1103/PhysRevB.15.2630} {\bibfield
  {journal} {\bibinfo  {journal} {Physical Review B}\ }\textbf {\bibinfo
  {volume} {15}},\ \bibinfo {pages} {2630} (\bibinfo {year}
  {1977})}\BibitemShut {NoStop}%
\bibitem [{\citenamefont {Degen}\ \emph {et~al.}(2009)\citenamefont {Degen},
  \citenamefont {Poggio}, \citenamefont {Mamin}, \citenamefont {Rettner},\ and\
  \citenamefont {Rugar}}]{DegenNanoscale2009a}%
  \BibitemOpen
  \bibfield  {author} {\bibinfo {author} {\bibfnamefont {C.~L.}\ \bibnamefont
  {Degen}}, \bibinfo {author} {\bibfnamefont {M.}~\bibnamefont {Poggio}},
  \bibinfo {author} {\bibfnamefont {H.~J.}\ \bibnamefont {Mamin}}, \bibinfo
  {author} {\bibfnamefont {C.~T.}\ \bibnamefont {Rettner}},\ and\ \bibinfo
  {author} {\bibfnamefont {D.}~\bibnamefont {Rugar}},\ }\href
  {https://doi.org/10.1073/pnas.0812068106} {\bibfield  {journal} {\bibinfo
  {journal} {Proceedings of the National Academy of Sciences}\ }\textbf
  {\bibinfo {volume} {106}},\ \bibinfo {pages} {1313} (\bibinfo {year}
  {2009})}\BibitemShut {NoStop}%
\bibitem [{\citenamefont {Huang}\ and\ \citenamefont
  {Maris}(2017)}]{huangEffective2017}%
  \BibitemOpen
  \bibfield  {author} {\bibinfo {author} {\bibfnamefont {Y.}~\bibnamefont
  {Huang}}\ and\ \bibinfo {author} {\bibfnamefont {H.~J.}\ \bibnamefont
  {Maris}},\ }\href {https://doi.org/10.1007/s10909-016-1669-7} {\bibfield
  {journal} {\bibinfo  {journal} {Journal of Low Temperature Physics}\ }\textbf
  {\bibinfo {volume} {186}},\ \bibinfo {pages} {208} (\bibinfo {year}
  {2017})}\BibitemShut {NoStop}%
\bibitem [{\citenamefont {McClintock}(1969)}]{McClintockField1969}%
  \BibitemOpen
  \bibfield  {author} {\bibinfo {author} {\bibfnamefont {P.}~\bibnamefont
  {McClintock}},\ }\href
  {https://doi.org/https://doi.org/10.1016/0375-9601(69)90517-9} {\bibfield
  {journal} {\bibinfo  {journal} {Physics Letters A}\ }\textbf {\bibinfo
  {volume} {29}},\ \bibinfo {pages} {453} (\bibinfo {year} {1969})}\BibitemShut
  {NoStop}%
\bibitem [{\citenamefont {Kashkanova}\ \emph {et~al.}(2017)\citenamefont
  {Kashkanova}, \citenamefont {Shkarin}, \citenamefont {Brown}, \citenamefont
  {{Flowers-Jacobs}}, \citenamefont {Childress}, \citenamefont {Hoch},
  \citenamefont {Hohmann}, \citenamefont {Ott}, \citenamefont {Reichel},\ and\
  \citenamefont {Harris}}]{KashkanovaSuperfluid2017a}%
  \BibitemOpen
  \bibfield  {author} {\bibinfo {author} {\bibfnamefont {A.~D.}\ \bibnamefont
  {Kashkanova}}, \bibinfo {author} {\bibfnamefont {A.~B.}\ \bibnamefont
  {Shkarin}}, \bibinfo {author} {\bibfnamefont {C.~D.}\ \bibnamefont {Brown}},
  \bibinfo {author} {\bibfnamefont {N.~E.}\ \bibnamefont {{Flowers-Jacobs}}},
  \bibinfo {author} {\bibfnamefont {L.}~\bibnamefont {Childress}}, \bibinfo
  {author} {\bibfnamefont {S.~W.}\ \bibnamefont {Hoch}}, \bibinfo {author}
  {\bibfnamefont {L.}~\bibnamefont {Hohmann}}, \bibinfo {author} {\bibfnamefont
  {K.}~\bibnamefont {Ott}}, \bibinfo {author} {\bibfnamefont {J.}~\bibnamefont
  {Reichel}},\ and\ \bibinfo {author} {\bibfnamefont {J.~G.~E.}\ \bibnamefont
  {Harris}},\ }\href {https://doi.org/10.1038/nphys3900} {\bibfield  {journal}
  {\bibinfo  {journal} {Nature Physics}\ }\textbf {\bibinfo {volume} {13}},\
  \bibinfo {pages} {74} (\bibinfo {year} {2017})}\BibitemShut {NoStop}%
\bibitem [{\citenamefont {Mamin}\ \emph {et~al.}(2003)\citenamefont {Mamin},
  \citenamefont {Budakian}, \citenamefont {Chui},\ and\ \citenamefont
  {Rugar}}]{MaminDetection2003}%
  \BibitemOpen
  \bibfield  {author} {\bibinfo {author} {\bibfnamefont {H.~J.}\ \bibnamefont
  {Mamin}}, \bibinfo {author} {\bibfnamefont {R.}~\bibnamefont {Budakian}},
  \bibinfo {author} {\bibfnamefont {B.~W.}\ \bibnamefont {Chui}},\ and\
  \bibinfo {author} {\bibfnamefont {D.}~\bibnamefont {Rugar}},\ }\href
  {https://doi.org/10.1103/PhysRevLett.91.207604} {\bibfield  {journal}
  {\bibinfo  {journal} {Physical Review Letters}\ }\textbf {\bibinfo {volume}
  {91}},\ \bibinfo {pages} {207604} (\bibinfo {year} {2003})}\BibitemShut
  {NoStop}%
\bibitem [{\citenamefont {Leirset}\ \emph {et~al.}(2013)\citenamefont
  {Leirset}, \citenamefont {Engan},\ and\ \citenamefont
  {Aksnes}}]{leirsetHeterodyne2013}%
  \BibitemOpen
  \bibfield  {author} {\bibinfo {author} {\bibfnamefont {E.}~\bibnamefont
  {Leirset}}, \bibinfo {author} {\bibfnamefont {H.~E.}\ \bibnamefont {Engan}},\
  and\ \bibinfo {author} {\bibfnamefont {A.}~\bibnamefont {Aksnes}},\ }\href
  {https://doi.org/10.1364/OE.21.019900} {\bibfield  {journal} {\bibinfo
  {journal} {Optics Express}\ }\textbf {\bibinfo {volume} {21}},\ \bibinfo
  {pages} {19900} (\bibinfo {year} {2013})}\BibitemShut {NoStop}%
\bibitem [{\citenamefont {Shkarin}\ \emph {et~al.}(2019)\citenamefont
  {Shkarin}, \citenamefont {Kashkanova}, \citenamefont {Brown}, \citenamefont
  {Garcia}, \citenamefont {Ott}, \citenamefont {Reichel},\ and\ \citenamefont
  {Harris}}]{ShkarinQuantum2019}%
  \BibitemOpen
  \bibfield  {author} {\bibinfo {author} {\bibfnamefont {A.~B.}\ \bibnamefont
  {Shkarin}}, \bibinfo {author} {\bibfnamefont {A.~D.}\ \bibnamefont
  {Kashkanova}}, \bibinfo {author} {\bibfnamefont {C.~D.}\ \bibnamefont
  {Brown}}, \bibinfo {author} {\bibfnamefont {S.}~\bibnamefont {Garcia}},
  \bibinfo {author} {\bibfnamefont {K.}~\bibnamefont {Ott}}, \bibinfo {author}
  {\bibfnamefont {J.}~\bibnamefont {Reichel}},\ and\ \bibinfo {author}
  {\bibfnamefont {J.~G.~E.}\ \bibnamefont {Harris}},\ }\href
  {https://doi.org/10.1103/PhysRevLett.122.153601} {\bibfield  {journal}
  {\bibinfo  {journal} {Physical Review Letters}\ }\textbf {\bibinfo {volume}
  {122}},\ \bibinfo {pages} {153601} (\bibinfo {year} {2019})}\BibitemShut
  {NoStop}%
\bibitem [{SI()}]{SI}%
  \BibitemOpen
  \href@noop {} {\bibinfo  {journal} {See supplemental material}\ }\BibitemShut
  {NoStop}%
\bibitem [{\citenamefont {Hunger}\ \emph {et~al.}(2012)\citenamefont {Hunger},
  \citenamefont {Deutsch}, \citenamefont {Barbour}, \citenamefont {Warburton},\
  and\ \citenamefont {Reichel}}]{Hunger2012}%
  \BibitemOpen
\bibfield  {journal} {  }\bibfield  {author} {\bibinfo {author} {\bibfnamefont
  {D.}~\bibnamefont {Hunger}}, \bibinfo {author} {\bibfnamefont
  {C.}~\bibnamefont {Deutsch}}, \bibinfo {author} {\bibfnamefont {R.~J.}\
  \bibnamefont {Barbour}}, \bibinfo {author} {\bibfnamefont {R.~J.}\
  \bibnamefont {Warburton}},\ and\ \bibinfo {author} {\bibfnamefont
  {J.}~\bibnamefont {Reichel}},\ }\href {https://doi.org/10.1063/1.3679721}
  {\bibfield  {journal} {\bibinfo  {journal} {AIP Advances}\ }\textbf {\bibinfo
  {volume} {2}},\ \bibinfo {pages} {012119} (\bibinfo {year}
  {2012})}\BibitemShut {NoStop}%
\bibitem [{\citenamefont {Yin}\ \emph {et~al.}(2011)\citenamefont {Yin},
  \citenamefont {Li}, \citenamefont {Chen}, \citenamefont {Li}, \citenamefont
  {Shung}, \citenamefont {Zhou}, \citenamefont {Jing}, \citenamefont {Mukai},
  \citenamefont {Mahon}, \citenamefont {Brenner}, \citenamefont {Edris},
  \citenamefont {Hoang},\ and\ \citenamefont {Narula}}]{Qifu2011}%
  \BibitemOpen
  \bibfield  {author} {\bibinfo {author} {\bibfnamefont {J.}~\bibnamefont
  {Yin}}, \bibinfo {author} {\bibfnamefont {J.}~\bibnamefont {Li}}, \bibinfo
  {author} {\bibfnamefont {Z.}~\bibnamefont {Chen}}, \bibinfo {author}
  {\bibfnamefont {X.}~\bibnamefont {Li}}, \bibinfo {author} {\bibfnamefont
  {K.~K.}\ \bibnamefont {Shung}}, \bibinfo {author} {\bibfnamefont
  {Q.}~\bibnamefont {Zhou}}, \bibinfo {author} {\bibfnamefont {J.}~\bibnamefont
  {Jing}}, \bibinfo {author} {\bibfnamefont {D.~S.}\ \bibnamefont {Mukai}},
  \bibinfo {author} {\bibfnamefont {S.~B.}\ \bibnamefont {Mahon}}, \bibinfo
  {author} {\bibfnamefont {M.}~\bibnamefont {Brenner}}, \bibinfo {author}
  {\bibfnamefont {A.}~\bibnamefont {Edris}}, \bibinfo {author} {\bibfnamefont
  {A.~K.~C.}\ \bibnamefont {Hoang}},\ and\ \bibinfo {author} {\bibfnamefont
  {J.}~\bibnamefont {Narula}},\ }\href {https://doi.org/10.1117/1.3589097}
  {\bibfield  {journal} {\bibinfo  {journal} {Journal of Biomedical Optics}\
  }\textbf {\bibinfo {volume} {16}},\ \bibinfo {pages} {060505} (\bibinfo
  {year} {2011})}\BibitemShut {NoStop}%
\bibitem [{\citenamefont {Millen}\ \emph {et~al.}(2020)\citenamefont {Millen},
  \citenamefont {Monteiro}, \citenamefont {Pettit},\ and\ \citenamefont
  {Vamivakas}}]{Millen2020}%
  \BibitemOpen
  \bibfield  {author} {\bibinfo {author} {\bibfnamefont {J.}~\bibnamefont
  {Millen}}, \bibinfo {author} {\bibfnamefont {T.~S.}\ \bibnamefont
  {Monteiro}}, \bibinfo {author} {\bibfnamefont {R.}~\bibnamefont {Pettit}},\
  and\ \bibinfo {author} {\bibfnamefont {A.~N.}\ \bibnamefont {Vamivakas}},\
  }\href {https://doi.org/10.1088/1361-6633/ab6100} {\bibfield  {journal}
  {\bibinfo  {journal} {Reports on Progress in Physics}\ }\textbf {\bibinfo
  {volume} {83}},\ \bibinfo {pages} {026401} (\bibinfo {year}
  {2020})}\BibitemShut {NoStop}%
\bibitem [{\citenamefont {Nimmrichter}\ \emph {et~al.}(2010)\citenamefont
  {Nimmrichter}, \citenamefont {Hammerer}, \citenamefont {Asenbaum},
  \citenamefont {Ritsch},\ and\ \citenamefont {Arndt}}]{NimmrichterMaster2010}%
  \BibitemOpen
  \bibfield  {author} {\bibinfo {author} {\bibfnamefont {S.}~\bibnamefont
  {Nimmrichter}}, \bibinfo {author} {\bibfnamefont {K.}~\bibnamefont
  {Hammerer}}, \bibinfo {author} {\bibfnamefont {P.}~\bibnamefont {Asenbaum}},
  \bibinfo {author} {\bibfnamefont {H.}~\bibnamefont {Ritsch}},\ and\ \bibinfo
  {author} {\bibfnamefont {M.}~\bibnamefont {Arndt}},\ }\href
  {https://doi.org/10.1088/1367-2630/12/8/083003} {\bibfield  {journal}
  {\bibinfo  {journal} {New Journal of Physics}\ }\textbf {\bibinfo {volume}
  {12}},\ \bibinfo {pages} {083003} (\bibinfo {year} {2010})}\BibitemShut
  {NoStop}%
\bibitem [{\citenamefont {Kiesel}\ \emph {et~al.}(2013)\citenamefont {Kiesel},
  \citenamefont {Blaser}, \citenamefont {Delic}, \citenamefont {Grass},
  \citenamefont {Kaltenbaek},\ and\ \citenamefont
  {Aspelmeyer}}]{KieselCavity2013a}%
  \BibitemOpen
  \bibfield  {author} {\bibinfo {author} {\bibfnamefont {N.}~\bibnamefont
  {Kiesel}}, \bibinfo {author} {\bibfnamefont {F.}~\bibnamefont {Blaser}},
  \bibinfo {author} {\bibfnamefont {U.}~\bibnamefont {Delic}}, \bibinfo
  {author} {\bibfnamefont {D.}~\bibnamefont {Grass}}, \bibinfo {author}
  {\bibfnamefont {R.}~\bibnamefont {Kaltenbaek}},\ and\ \bibinfo {author}
  {\bibfnamefont {M.}~\bibnamefont {Aspelmeyer}},\ }\href
  {https://doi.org/10.1073/pnas.1309167110} {\bibfield  {journal} {\bibinfo
  {journal} {Proceedings of the National Academy of Sciences}\ }\textbf
  {\bibinfo {volume} {110}},\ \bibinfo {pages} {14180} (\bibinfo {year}
  {2013})}\BibitemShut {NoStop}%
\bibitem [{\citenamefont {Aspelmeyer}\ \emph {et~al.}(2014)\citenamefont
  {Aspelmeyer}, \citenamefont {Kippenberg},\ and\ \citenamefont
  {Marquardt}}]{aspelmeyer_cavity-opto_review}%
  \BibitemOpen
  \bibfield  {author} {\bibinfo {author} {\bibfnamefont {M.}~\bibnamefont
  {Aspelmeyer}}, \bibinfo {author} {\bibfnamefont {T.~J.}\ \bibnamefont
  {Kippenberg}},\ and\ \bibinfo {author} {\bibfnamefont {F.}~\bibnamefont
  {Marquardt}},\ }\href {https://doi.org/10.1103/RevModPhys.86.1391} {\bibfield
   {journal} {\bibinfo  {journal} {Reviews of Modern Physics}\ }\textbf
  {\bibinfo {volume} {86}},\ \bibinfo {pages} {1391} (\bibinfo {year}
  {2014})}\BibitemShut {NoStop}%
\bibitem [{\citenamefont {Baym}\ \emph {et~al.}(1969)\citenamefont {Baym},
  \citenamefont {Barrera},\ and\ \citenamefont {Pethick}}]{BaymMobility1969}%
  \BibitemOpen
  \bibfield  {author} {\bibinfo {author} {\bibfnamefont {G.}~\bibnamefont
  {Baym}}, \bibinfo {author} {\bibfnamefont {R.~G.}\ \bibnamefont {Barrera}},\
  and\ \bibinfo {author} {\bibfnamefont {C.~J.}\ \bibnamefont {Pethick}},\
  }\href {https://doi.org/10.1103/PhysRevLett.22.20} {\bibfield  {journal}
  {\bibinfo  {journal} {Physical Review Letters}\ }\textbf {\bibinfo {volume}
  {22}},\ \bibinfo {pages} {20} (\bibinfo {year} {1969})}\BibitemShut {NoStop}%
\bibitem [{\citenamefont {Kramer}(1970)}]{KramerLowTemperature1970}%
  \BibitemOpen
  \bibfield  {author} {\bibinfo {author} {\bibfnamefont {L.}~\bibnamefont
  {Kramer}},\ }\href {https://doi.org/10.1103/PhysRevA.1.1517} {\bibfield
  {journal} {\bibinfo  {journal} {Physical Review A}\ }\textbf {\bibinfo
  {volume} {1}},\ \bibinfo {pages} {1517} (\bibinfo {year} {1970})}\BibitemShut
  {NoStop}%
\bibitem [{\citenamefont {Borghesani}(2007)}]{BorghesaniIons2007}%
  \BibitemOpen
  \bibfield  {author} {\bibinfo {author} {\bibfnamefont {A.~F.}\ \bibnamefont
  {Borghesani}},\ }\href {https://academic.oup.com/book/4252} {\emph {\bibinfo
  {title} {Ions and Electrons in Liquid Helium}}},\ \bibinfo {series}
  {International Series of Monographs on Physics}\ No.\ \bibinfo {number}
  {137}\ (\bibinfo  {publisher} {{Oxford University Press}},\ \bibinfo
  {address} {{Oxford ; New York}},\ \bibinfo {year} {2007})\BibitemShut
  {NoStop}%
\bibitem [{\citenamefont {Budakian}\ \emph {et~al.}(2005)\citenamefont
  {Budakian}, \citenamefont {Mamin}, \citenamefont {Chui},\ and\ \citenamefont
  {Rugar}}]{RugarFluc}%
  \BibitemOpen
  \bibfield  {author} {\bibinfo {author} {\bibfnamefont {R.}~\bibnamefont
  {Budakian}}, \bibinfo {author} {\bibfnamefont {H.~J.}\ \bibnamefont {Mamin}},
  \bibinfo {author} {\bibfnamefont {B.~W.}\ \bibnamefont {Chui}},\ and\
  \bibinfo {author} {\bibfnamefont {D.}~\bibnamefont {Rugar}},\ }\href
  {https://doi.org/10.1126/science.1106718} {\bibfield  {journal} {\bibinfo
  {journal} {Science}\ }\textbf {\bibinfo {volume} {307}},\ \bibinfo {pages}
  {408} (\bibinfo {year} {2005})},\ \Eprint
  {https://arxiv.org/abs/https://www.science.org/doi/pdf/10.1126/science.1106718}
  {https://www.science.org/doi/pdf/10.1126/science.1106718} \BibitemShut
  {NoStop}%
\bibitem [{\citenamefont {Grob}\ \emph {et~al.}(2019)\citenamefont {Grob},
  \citenamefont {Krass}, \citenamefont {Héritier}, \citenamefont {Pachlatko},
  \citenamefont {Rhensius}, \citenamefont {Košata}, \citenamefont {Moores},
  \citenamefont {Takahashi}, \citenamefont {Eichler},\ and\ \citenamefont
  {Degen}}]{Grob2019}%
  \BibitemOpen
  \bibfield  {author} {\bibinfo {author} {\bibfnamefont {U.}~\bibnamefont
  {Grob}}, \bibinfo {author} {\bibfnamefont {M.~D.}\ \bibnamefont {Krass}},
  \bibinfo {author} {\bibfnamefont {M.}~\bibnamefont {Héritier}}, \bibinfo
  {author} {\bibfnamefont {R.}~\bibnamefont {Pachlatko}}, \bibinfo {author}
  {\bibfnamefont {J.}~\bibnamefont {Rhensius}}, \bibinfo {author}
  {\bibfnamefont {J.}~\bibnamefont {Košata}}, \bibinfo {author} {\bibfnamefont
  {B.~A.}\ \bibnamefont {Moores}}, \bibinfo {author} {\bibfnamefont
  {H.}~\bibnamefont {Takahashi}}, \bibinfo {author} {\bibfnamefont
  {A.}~\bibnamefont {Eichler}},\ and\ \bibinfo {author} {\bibfnamefont {C.~L.}\
  \bibnamefont {Degen}},\ }\href {https://doi.org/10.1021/acs.nanolett.9b03048}
  {\bibfield  {journal} {\bibinfo  {journal} {Nano Letters}\ }\textbf {\bibinfo
  {volume} {19}},\ \bibinfo {pages} {7935} (\bibinfo {year}
  {2019})}\BibitemShut {NoStop}%
\bibitem [{\citenamefont {M\o{}lmer}\ and\ \citenamefont
  {S\o{}rensen}(1999)}]{Molmer1999}%
  \BibitemOpen
  \bibfield  {author} {\bibinfo {author} {\bibfnamefont {K.}~\bibnamefont
  {M\o{}lmer}}\ and\ \bibinfo {author} {\bibfnamefont {A.}~\bibnamefont
  {S\o{}rensen}},\ }\href {https://doi.org/10.1103/PhysRevLett.82.1835}
  {\bibfield  {journal} {\bibinfo  {journal} {Phys. Rev. Lett.}\ }\textbf
  {\bibinfo {volume} {82}},\ \bibinfo {pages} {1835} (\bibinfo {year}
  {1999})}\BibitemShut {NoStop}%
\bibitem [{\citenamefont {Volodin}\ \emph {et~al.}(1977)\citenamefont
  {Volodin}, \citenamefont {Kha\u{i}kin},\ and\ \citenamefont
  {\'{E}del'man}}]{Volodin1977}%
  \BibitemOpen
  \bibfield  {author} {\bibinfo {author} {\bibfnamefont {A.~P.}\ \bibnamefont
  {Volodin}}, \bibinfo {author} {\bibfnamefont {M.~S.}\ \bibnamefont
  {Kha\u{i}kin}},\ and\ \bibinfo {author} {\bibfnamefont {V.~S.}\ \bibnamefont
  {\'{E}del'man}},\ }\href {http://jetpletters.ru/ps/1384/article_20986.pdf}
  {\bibfield  {journal} {\bibinfo  {journal} {Journal of Experimental and
  Theoretical Physics Letters}\ }\textbf {\bibinfo {volume} {26}},\ \bibinfo
  {pages} {543} (\bibinfo {year} {1977})}\BibitemShut {NoStop}%
\bibitem [{\citenamefont {Yadav}\ \emph {et~al.}(2021)\citenamefont {Yadav},
  \citenamefont {Sen},\ and\ \citenamefont {Ghosh}}]{Yadav2021}%
  \BibitemOpen
  \bibfield  {author} {\bibinfo {author} {\bibfnamefont {N.}~\bibnamefont
  {Yadav}}, \bibinfo {author} {\bibfnamefont {P.}~\bibnamefont {Sen}},\ and\
  \bibinfo {author} {\bibfnamefont {A.}~\bibnamefont {Ghosh}},\ }\href
  {https://doi.org/10.1126/sciadv.abi7128} {\bibfield  {journal} {\bibinfo
  {journal} {Science Advances}\ }\textbf {\bibinfo {volume} {7}},\ \bibinfo
  {pages} {eabi7128} (\bibinfo {year} {2021})},\ \Eprint
  {https://arxiv.org/abs/https://www.science.org/doi/pdf/10.1126/sciadv.abi7128}
  {https://www.science.org/doi/pdf/10.1126/sciadv.abi7128} \BibitemShut
  {NoStop}%
\bibitem [{\citenamefont {Slenczka}\ and\ \citenamefont
  {Toennies}(2022)}]{slenczka2022molecules}%
  \BibitemOpen
  \bibfield  {author} {\bibinfo {author} {\bibfnamefont {A.}~\bibnamefont
  {Slenczka}}\ and\ \bibinfo {author} {\bibfnamefont {J.}~\bibnamefont
  {Toennies}},\ }\href {https://books.google.com/books?id=nOlxEAAAQBAJ} {\emph
  {\bibinfo {title} {Molecules in Superfluid Helium Nanodroplets: Spectroscopy,
  Structure, and Dynamics}}},\ Topics in Applied Physics\ (\bibinfo
  {publisher} {Springer International Publishing},\ \bibinfo {year}
  {2022})\BibitemShut {NoStop}%
\bibitem [{\citenamefont {Toennies}\ and\ \citenamefont
  {Vilesov}(2004)}]{Toennies2004}%
  \BibitemOpen
  \bibfield  {author} {\bibinfo {author} {\bibfnamefont {J.~P.}\ \bibnamefont
  {Toennies}}\ and\ \bibinfo {author} {\bibfnamefont {A.~F.}\ \bibnamefont
  {Vilesov}},\ }\href {https://doi.org/https://doi.org/10.1002/anie.200300611}
  {\bibfield  {journal} {\bibinfo  {journal} {Angewandte Chemie International
  Edition}\ }\textbf {\bibinfo {volume} {43}},\ \bibinfo {pages} {2622}
  (\bibinfo {year} {2004})},\ \Eprint
  {https://arxiv.org/abs/https://onlinelibrary.wiley.com/doi/pdf/10.1002/anie.200300611}
  {https://onlinelibrary.wiley.com/doi/pdf/10.1002/anie.200300611} \BibitemShut
  {NoStop}%
\end{thebibliography}%


\begin{thebibliography}{24}%
\makeatletter
\providecommand \@ifxundefined [1]{%
 \@ifx{#1\undefined}
}%
\providecommand \@ifnum [1]{%
 \ifnum #1\expandafter \@firstoftwo
 \else \expandafter \@secondoftwo
 \fi
}%
\providecommand \@ifx [1]{%
 \ifx #1\expandafter \@firstoftwo
 \else \expandafter \@secondoftwo
 \fi
}%
\providecommand \natexlab [1]{#1}%
\providecommand \enquote  [1]{``#1''}%
\providecommand \bibnamefont  [1]{#1}%
\providecommand \bibfnamefont [1]{#1}%
\providecommand \citenamefont [1]{#1}%
\providecommand \href@noop [0]{\@secondoftwo}%
\providecommand \href [0]{\begingroup \@sanitize@url \@href}%
\providecommand \@href[1]{\@@startlink{#1}\@@href}%
\providecommand \@@href[1]{\endgroup#1\@@endlink}%
\providecommand \@sanitize@url [0]{\catcode `\\12\catcode `\$12\catcode
  `\&12\catcode `\#12\catcode `\^12\catcode `\_12\catcode `\%12\relax}%
\providecommand \@@startlink[1]{}%
\providecommand \@@endlink[0]{}%
\providecommand \url  [0]{\begingroup\@sanitize@url \@url }%
\providecommand \@url [1]{\endgroup\@href {#1}{\urlprefix }}%
\providecommand \urlprefix  [0]{URL }%
\providecommand \Eprint [0]{\href }%
\providecommand \doibase [0]{https://doi.org/}%
\providecommand \selectlanguage [0]{\@gobble}%
\providecommand \bibinfo  [0]{\@secondoftwo}%
\providecommand \bibfield  [0]{\@secondoftwo}%
\providecommand \translation [1]{[#1]}%
\providecommand \BibitemOpen [0]{}%
\providecommand \bibitemStop [0]{}%
\providecommand \bibitemNoStop [0]{.\EOS\space}%
\providecommand \EOS [0]{\spacefactor3000\relax}%
\providecommand \BibitemShut  [1]{\csname bibitem#1\endcsname}%
\let\auto@bib@innerbib\@empty
\bibitem [{\citenamefont {Grimes}\ and\ \citenamefont
  {Adams}(1990)}]{grimesInfrared1990}%
  \BibitemOpen
  \bibfield  {author} {\bibinfo {author} {\bibfnamefont {C.~C.}\ \bibnamefont
  {Grimes}}\ and\ \bibinfo {author} {\bibfnamefont {G.}~\bibnamefont {Adams}},\
  }\href {https://doi.org/10.1103/PhysRevB.41.6366} {\bibfield  {journal}
  {\bibinfo  {journal} {Physical Review B}\ }\textbf {\bibinfo {volume} {41}},\
  \bibinfo {pages} {6366} (\bibinfo {year} {1990})}\BibitemShut {NoStop}%
\bibitem [{\citenamefont {Classen}\ \emph {et~al.}(1998)\citenamefont
  {Classen}, \citenamefont {Su}, \citenamefont {Mohazzab},\ and\ \citenamefont
  {Maris}}]{classenElectrons1998}%
  \BibitemOpen
  \bibfield  {author} {\bibinfo {author} {\bibfnamefont {J.}~\bibnamefont
  {Classen}}, \bibinfo {author} {\bibfnamefont {C.-K.}\ \bibnamefont {Su}},
  \bibinfo {author} {\bibfnamefont {M.}~\bibnamefont {Mohazzab}},\ and\
  \bibinfo {author} {\bibfnamefont {H.~J.}\ \bibnamefont {Maris}},\ }\href
  {https://doi.org/10.1103/PhysRevB.57.3000} {\bibfield  {journal} {\bibinfo
  {journal} {Physical Review B}\ }\textbf {\bibinfo {volume} {57}},\ \bibinfo
  {pages} {3000} (\bibinfo {year} {1998})}\BibitemShut {NoStop}%
\bibitem [{\citenamefont {Konstantinov}\ and\ \citenamefont
  {Maris}(2003)}]{konstantinovDetection2003}%
  \BibitemOpen
  \bibfield  {author} {\bibinfo {author} {\bibfnamefont {D.}~\bibnamefont
  {Konstantinov}}\ and\ \bibinfo {author} {\bibfnamefont {H.~J.}\ \bibnamefont
  {Maris}},\ }\href {https://doi.org/10.1103/PhysRevLett.90.025302} {\bibfield
  {journal} {\bibinfo  {journal} {Physical Review Letters}\ }\textbf {\bibinfo
  {volume} {90}},\ \bibinfo {pages} {025302} (\bibinfo {year}
  {2003})}\BibitemShut {NoStop}%
\bibitem [{\citenamefont {Xing}\ and\ \citenamefont
  {Maris}(2020)}]{xingElectrons2020}%
  \BibitemOpen
  \bibfield  {author} {\bibinfo {author} {\bibfnamefont {Y.}~\bibnamefont
  {Xing}}\ and\ \bibinfo {author} {\bibfnamefont {H.~J.}\ \bibnamefont
  {Maris}},\ }\href {https://doi.org/10.1007/s10909-020-02422-5} {\bibfield
  {journal} {\bibinfo  {journal} {Journal of Low Temperature Physics}\ }\textbf
  {\bibinfo {volume} {201}},\ \bibinfo {pages} {634} (\bibinfo {year}
  {2020})}\BibitemShut {NoStop}%
\bibitem [{\citenamefont {Roche}\ \emph {et~al.}(1997)\citenamefont {Roche},
  \citenamefont {Deville}, \citenamefont {Appleyard},\ and\ \citenamefont
  {Williams}}]{rocheMeasurement1997}%
  \BibitemOpen
  \bibfield  {author} {\bibinfo {author} {\bibfnamefont {P.}~\bibnamefont
  {Roche}}, \bibinfo {author} {\bibfnamefont {G.}~\bibnamefont {Deville}},
  \bibinfo {author} {\bibfnamefont {N.~J.}\ \bibnamefont {Appleyard}},\ and\
  \bibinfo {author} {\bibfnamefont {F.~I.~B.}\ \bibnamefont {Williams}},\
  }\href {https://doi.org/10.1007/BF02395922} {\bibfield  {journal} {\bibinfo
  {journal} {Journal of Low Temperature Physics}\ }\textbf {\bibinfo {volume}
  {106}},\ \bibinfo {pages} {565} (\bibinfo {year} {1997})}\BibitemShut
  {NoStop}%
\bibitem [{\citenamefont {Bruus}(2012)}]{BruusAcoustofluidics2012a}%
  \BibitemOpen
  \bibfield  {author} {\bibinfo {author} {\bibfnamefont {H.}~\bibnamefont
  {Bruus}},\ }\href {https://doi.org/10.1039/c2lc21068a} {\bibfield  {journal}
  {\bibinfo  {journal} {Lab on a Chip}\ }\textbf {\bibinfo {volume} {12}},\
  \bibinfo {pages} {1014} (\bibinfo {year} {2012})}\BibitemShut {NoStop}%
\bibitem [{\citenamefont {Huang}\ and\ \citenamefont
  {Maris}(2017)}]{huangEffective2017}%
  \BibitemOpen
  \bibfield  {author} {\bibinfo {author} {\bibfnamefont {Y.}~\bibnamefont
  {Huang}}\ and\ \bibinfo {author} {\bibfnamefont {H.~J.}\ \bibnamefont
  {Maris}},\ }\href {https://doi.org/10.1007/s10909-016-1669-7} {\bibfield
  {journal} {\bibinfo  {journal} {Journal of Low Temperature Physics}\ }\textbf
  {\bibinfo {volume} {186}},\ \bibinfo {pages} {208} (\bibinfo {year}
  {2017})}\BibitemShut {NoStop}%
\bibitem [{\citenamefont {Han}\ and\ \citenamefont
  {Lee}(2014)}]{HanSpecification2014a}%
  \BibitemOpen
  \bibfield  {author} {\bibinfo {author} {\bibfnamefont {H.-S.}\ \bibnamefont
  {Han}}\ and\ \bibinfo {author} {\bibfnamefont {K.-H.}\ \bibnamefont {Lee}},\
  }\href {https://doi.org/10.1007/s12206-014-0803-1} {\bibfield  {journal}
  {\bibinfo  {journal} {Journal of Mechanical Science and Technology}\ }\textbf
  {\bibinfo {volume} {28}},\ \bibinfo {pages} {3425} (\bibinfo {year}
  {2014})}\BibitemShut {NoStop}%
\bibitem [{\citenamefont {Shkarin}\ \emph {et~al.}(2019)\citenamefont
  {Shkarin}, \citenamefont {Kashkanova}, \citenamefont {Brown}, \citenamefont
  {Garcia}, \citenamefont {Ott}, \citenamefont {Reichel},\ and\ \citenamefont
  {Harris}}]{shkarinQuantum2019a}%
  \BibitemOpen
  \bibfield  {author} {\bibinfo {author} {\bibfnamefont {A.~B.}\ \bibnamefont
  {Shkarin}}, \bibinfo {author} {\bibfnamefont {A.~D.}\ \bibnamefont
  {Kashkanova}}, \bibinfo {author} {\bibfnamefont {C.~D.}\ \bibnamefont
  {Brown}}, \bibinfo {author} {\bibfnamefont {S.}~\bibnamefont {Garcia}},
  \bibinfo {author} {\bibfnamefont {K.}~\bibnamefont {Ott}}, \bibinfo {author}
  {\bibfnamefont {J.}~\bibnamefont {Reichel}},\ and\ \bibinfo {author}
  {\bibfnamefont {J.~G.~E.}\ \bibnamefont {Harris}},\ }\bibfield  {journal}
  {\bibinfo  {journal} {Physical Review Letters}\ }\textbf {\bibinfo {volume}
  {122}},\ \href {https://doi.org/10.1103/PhysRevLett.122.153601}
  {10.1103/PhysRevLett.122.153601} (\bibinfo {year} {2019})\BibitemShut
  {NoStop}%
\bibitem [{\citenamefont {Kashkanova}\ \emph {et~al.}(2017)\citenamefont
  {Kashkanova}, \citenamefont {Shkarin}, \citenamefont {Brown}, \citenamefont
  {{Flowers-Jacobs}}, \citenamefont {Childress}, \citenamefont {Hoch},
  \citenamefont {Hohmann}, \citenamefont {Ott}, \citenamefont {Reichel},\ and\
  \citenamefont {Harris}}]{KashkanovaSuperfluid2017a}%
  \BibitemOpen
  \bibfield  {author} {\bibinfo {author} {\bibfnamefont {A.~D.}\ \bibnamefont
  {Kashkanova}}, \bibinfo {author} {\bibfnamefont {A.~B.}\ \bibnamefont
  {Shkarin}}, \bibinfo {author} {\bibfnamefont {C.~D.}\ \bibnamefont {Brown}},
  \bibinfo {author} {\bibfnamefont {N.~E.}\ \bibnamefont {{Flowers-Jacobs}}},
  \bibinfo {author} {\bibfnamefont {L.}~\bibnamefont {Childress}}, \bibinfo
  {author} {\bibfnamefont {S.~W.}\ \bibnamefont {Hoch}}, \bibinfo {author}
  {\bibfnamefont {L.}~\bibnamefont {Hohmann}}, \bibinfo {author} {\bibfnamefont
  {K.}~\bibnamefont {Ott}}, \bibinfo {author} {\bibfnamefont {J.}~\bibnamefont
  {Reichel}},\ and\ \bibinfo {author} {\bibfnamefont {J.~G.~E.}\ \bibnamefont
  {Harris}},\ }\href {https://doi.org/10.1038/nphys3900} {\bibfield  {journal}
  {\bibinfo  {journal} {Nature Physics}\ }\textbf {\bibinfo {volume} {13}},\
  \bibinfo {pages} {74} (\bibinfo {year} {2017})}\BibitemShut {NoStop}%
\bibitem [{\citenamefont {Leirset}\ \emph {et~al.}(2013)\citenamefont
  {Leirset}, \citenamefont {Engan},\ and\ \citenamefont
  {Aksnes}}]{leirsetHeterodyne2013}%
  \BibitemOpen
  \bibfield  {author} {\bibinfo {author} {\bibfnamefont {E.}~\bibnamefont
  {Leirset}}, \bibinfo {author} {\bibfnamefont {H.~E.}\ \bibnamefont {Engan}},\
  and\ \bibinfo {author} {\bibfnamefont {A.}~\bibnamefont {Aksnes}},\ }\href
  {https://doi.org/10.1364/OE.21.019900} {\bibfield  {journal} {\bibinfo
  {journal} {Optics Express}\ }\textbf {\bibinfo {volume} {21}},\ \bibinfo
  {pages} {19900} (\bibinfo {year} {2013})}\BibitemShut {NoStop}%
\bibitem [{\citenamefont {Shkarin}(2018)}]{shkarinQuantuma}%
  \BibitemOpen
  \bibfield  {author} {\bibinfo {author} {\bibfnamefont {A.}~\bibnamefont
  {Shkarin}},\ }\emph {\bibinfo {title} {Quantum {{Optomechanics}} with
  {{Superfluid Helium}}}},\ \href@noop {} {Ph.D. thesis},\ \bibinfo  {school}
  {Yale University} (\bibinfo {year} {2018})\BibitemShut {NoStop}%
\bibitem [{\citenamefont {Kashkanova}(2017)}]{kashkanovaOptomechanics}%
  \BibitemOpen
  \bibfield  {author} {\bibinfo {author} {\bibfnamefont {A.~D.}\ \bibnamefont
  {Kashkanova}},\ }\emph {\bibinfo {title} {Optomechanics with {{Superfluid
  Helium}}}},\ \href@noop {} {Ph.D. thesis},\ \bibinfo  {school} {Yale
  University} (\bibinfo {year} {2017})\BibitemShut {NoStop}%
\bibitem [{\citenamefont {Kiesel}\ \emph {et~al.}(2013)\citenamefont {Kiesel},
  \citenamefont {Blaser}, \citenamefont {Delic}, \citenamefont {Grass},
  \citenamefont {Kaltenbaek},\ and\ \citenamefont
  {Aspelmeyer}}]{KieselCavity2013a}%
  \BibitemOpen
  \bibfield  {author} {\bibinfo {author} {\bibfnamefont {N.}~\bibnamefont
  {Kiesel}}, \bibinfo {author} {\bibfnamefont {F.}~\bibnamefont {Blaser}},
  \bibinfo {author} {\bibfnamefont {U.}~\bibnamefont {Delic}}, \bibinfo
  {author} {\bibfnamefont {D.}~\bibnamefont {Grass}}, \bibinfo {author}
  {\bibfnamefont {R.}~\bibnamefont {Kaltenbaek}},\ and\ \bibinfo {author}
  {\bibfnamefont {M.}~\bibnamefont {Aspelmeyer}},\ }\href
  {https://doi.org/10.1073/pnas.1309167110} {\bibfield  {journal} {\bibinfo
  {journal} {Proceedings of the National Academy of Sciences}\ }\textbf
  {\bibinfo {volume} {110}},\ \bibinfo {pages} {14180} (\bibinfo {year}
  {2013})}\BibitemShut {NoStop}%
\bibitem [{\citenamefont {Nimmrichter}\ \emph {et~al.}(2010)\citenamefont
  {Nimmrichter}, \citenamefont {Hammerer}, \citenamefont {Asenbaum},
  \citenamefont {Ritsch},\ and\ \citenamefont
  {Arndt}}]{NimmrichterMaster2010a}%
  \BibitemOpen
  \bibfield  {author} {\bibinfo {author} {\bibfnamefont {S.}~\bibnamefont
  {Nimmrichter}}, \bibinfo {author} {\bibfnamefont {K.}~\bibnamefont
  {Hammerer}}, \bibinfo {author} {\bibfnamefont {P.}~\bibnamefont {Asenbaum}},
  \bibinfo {author} {\bibfnamefont {H.}~\bibnamefont {Ritsch}},\ and\ \bibinfo
  {author} {\bibfnamefont {M.}~\bibnamefont {Arndt}},\ }\href
  {https://doi.org/10.1088/1367-2630/12/8/083003} {\bibfield  {journal}
  {\bibinfo  {journal} {New Journal of Physics}\ }\textbf {\bibinfo {volume}
  {12}},\ \bibinfo {pages} {083003} (\bibinfo {year} {2010})}\BibitemShut
  {NoStop}%
\bibitem [{\citenamefont {Allum}\ \emph {et~al.}(1977)\citenamefont {Allum},
  \citenamefont {McClintock}, \citenamefont {Phillips}, \citenamefont
  {Bowley},\ and\ \citenamefont {Vinen}}]{Allum1977}%
  \BibitemOpen
  \bibfield  {author} {\bibinfo {author} {\bibfnamefont {D.~R.}\ \bibnamefont
  {Allum}}, \bibinfo {author} {\bibfnamefont {P.~V.~E.}\ \bibnamefont
  {McClintock}}, \bibinfo {author} {\bibfnamefont {A.}~\bibnamefont
  {Phillips}}, \bibinfo {author} {\bibfnamefont {R.~M.}\ \bibnamefont
  {Bowley}},\ and\ \bibinfo {author} {\bibfnamefont {W.~F.}\ \bibnamefont
  {Vinen}},\ }\href {https://doi.org/10.1098/rsta.1977.0008} {\bibfield
  {journal} {\bibinfo  {journal} {Philosophical Transactions of the Royal
  Society of London. Series A, Mathematical and Physical Sciences}\ }\textbf
  {\bibinfo {volume} {284}},\ \bibinfo {pages} {179} (\bibinfo {year}
  {1977})}\BibitemShut {NoStop}%
\bibitem [{\citenamefont {Nancolas}\ \emph {et~al.}(1985)\citenamefont
  {Nancolas}, \citenamefont {Ellis}, \citenamefont {McClintock},\ and\
  \citenamefont {Bowley}}]{Nancolas1985}%
  \BibitemOpen
  \bibfield  {author} {\bibinfo {author} {\bibfnamefont {G.~G.}\ \bibnamefont
  {Nancolas}}, \bibinfo {author} {\bibfnamefont {T.}~\bibnamefont {Ellis}},
  \bibinfo {author} {\bibfnamefont {P.~V.~E.}\ \bibnamefont {McClintock}},\
  and\ \bibinfo {author} {\bibfnamefont {R.~M.}\ \bibnamefont {Bowley}},\
  }\href {https://doi.org/10.1038/316797a0} {\bibfield  {journal} {\bibinfo
  {journal} {Nature}\ }\textbf {\bibinfo {volume} {316}},\ \bibinfo {pages}
  {797} (\bibinfo {year} {1985})}\BibitemShut {NoStop}%
\bibitem [{\citenamefont {Baym}\ \emph {et~al.}(1969)\citenamefont {Baym},
  \citenamefont {Barrera},\ and\ \citenamefont {Pethick}}]{BaymMobility1969}%
  \BibitemOpen
  \bibfield  {author} {\bibinfo {author} {\bibfnamefont {G.}~\bibnamefont
  {Baym}}, \bibinfo {author} {\bibfnamefont {R.~G.}\ \bibnamefont {Barrera}},\
  and\ \bibinfo {author} {\bibfnamefont {C.~J.}\ \bibnamefont {Pethick}},\
  }\href {https://doi.org/10.1103/PhysRevLett.22.20} {\bibfield  {journal}
  {\bibinfo  {journal} {Physical Review Letters}\ }\textbf {\bibinfo {volume}
  {22}},\ \bibinfo {pages} {20} (\bibinfo {year} {1969})}\BibitemShut {NoStop}%
\bibitem [{\citenamefont {Schwarz}(1972)}]{schwarzHe1972}%
  \BibitemOpen
  \bibfield  {author} {\bibinfo {author} {\bibfnamefont {K.~W.}\ \bibnamefont
  {Schwarz}},\ }\href {https://doi.org/10.1103/PhysRevA.6.1947} {\bibfield
  {journal} {\bibinfo  {journal} {Physical Review A}\ }\textbf {\bibinfo
  {volume} {6}},\ \bibinfo {pages} {1947} (\bibinfo {year} {1972})}\BibitemShut
  {NoStop}%
\bibitem [{\citenamefont {Mamin}\ \emph
  {et~al.}(2003{\natexlab{a}})\citenamefont {Mamin}, \citenamefont {Budakian},
  \citenamefont {Chui},\ and\ \citenamefont {Rugar}}]{MaminDetection2003}%
  \BibitemOpen
  \bibfield  {author} {\bibinfo {author} {\bibfnamefont {H.~J.}\ \bibnamefont
  {Mamin}}, \bibinfo {author} {\bibfnamefont {R.}~\bibnamefont {Budakian}},
  \bibinfo {author} {\bibfnamefont {B.~W.}\ \bibnamefont {Chui}},\ and\
  \bibinfo {author} {\bibfnamefont {D.}~\bibnamefont {Rugar}},\ }\bibfield
  {journal} {\bibinfo  {journal} {Physical Review Letters}\ }\textbf {\bibinfo
  {volume} {91}},\ \href {https://doi.org/10.1103/PhysRevLett.91.207604}
  {10.1103/PhysRevLett.91.207604} (\bibinfo {year}
  {2003}{\natexlab{a}})\BibitemShut {NoStop}%
\bibitem [{\citenamefont {Rugar}\ \emph {et~al.}(2004)\citenamefont {Rugar},
  \citenamefont {Budakian}, \citenamefont {Mamin},\ and\ \citenamefont
  {Chui}}]{RugarSingle2004}%
  \BibitemOpen
  \bibfield  {author} {\bibinfo {author} {\bibfnamefont {D.}~\bibnamefont
  {Rugar}}, \bibinfo {author} {\bibfnamefont {R.}~\bibnamefont {Budakian}},
  \bibinfo {author} {\bibfnamefont {H.~J.}\ \bibnamefont {Mamin}},\ and\
  \bibinfo {author} {\bibfnamefont {B.~W.}\ \bibnamefont {Chui}},\ }\href
  {https://doi.org/10.1038/nature02658} {\bibfield  {journal} {\bibinfo
  {journal} {Nature}\ }\textbf {\bibinfo {volume} {430}},\ \bibinfo {pages}
  {329} (\bibinfo {year} {2004})}\BibitemShut {NoStop}%
\bibitem [{\citenamefont {Mamin}\ \emph
  {et~al.}(2003{\natexlab{b}})\citenamefont {Mamin}, \citenamefont {Budakian},
  \citenamefont {Chui},\ and\ \citenamefont {Rugar}}]{MaminDetection2003a}%
  \BibitemOpen
  \bibfield  {author} {\bibinfo {author} {\bibfnamefont {H.~J.}\ \bibnamefont
  {Mamin}}, \bibinfo {author} {\bibfnamefont {R.}~\bibnamefont {Budakian}},
  \bibinfo {author} {\bibfnamefont {B.~W.}\ \bibnamefont {Chui}},\ and\
  \bibinfo {author} {\bibfnamefont {D.}~\bibnamefont {Rugar}},\ }\bibfield
  {journal} {\bibinfo  {journal} {Physical Review Letters}\ }\textbf {\bibinfo
  {volume} {91}},\ \href {https://doi.org/10.1103/PhysRevLett.91.207604}
  {10.1103/PhysRevLett.91.207604} (\bibinfo {year}
  {2003}{\natexlab{b}})\BibitemShut {NoStop}%
\bibitem [{\citenamefont {Ma}\ \emph {et~al.}(2015)\citenamefont {Ma},
  \citenamefont {You}, \citenamefont {Si}, \citenamefont {Xiong}, \citenamefont
  {Li}, \citenamefont {Yang},\ and\ \citenamefont
  {Wu}}]{ma_OMIT-mechanical-drive_opto}%
  \BibitemOpen
  \bibfield  {author} {\bibinfo {author} {\bibfnamefont {J.}~\bibnamefont
  {Ma}}, \bibinfo {author} {\bibfnamefont {C.}~\bibnamefont {You}}, \bibinfo
  {author} {\bibfnamefont {L.-G.}\ \bibnamefont {Si}}, \bibinfo {author}
  {\bibfnamefont {H.}~\bibnamefont {Xiong}}, \bibinfo {author} {\bibfnamefont
  {J.}~\bibnamefont {Li}}, \bibinfo {author} {\bibfnamefont {X.}~\bibnamefont
  {Yang}},\ and\ \bibinfo {author} {\bibfnamefont {Y.}~\bibnamefont {Wu}},\
  }\href {https://doi.org/10.1038/srep11278} {\bibfield  {journal} {\bibinfo
  {journal} {Scientific Reports}\ }\textbf {\bibinfo {volume} {5}},\ \bibinfo
  {pages} {11278} (\bibinfo {year} {2015})}\BibitemShut {NoStop}%
\bibitem [{\citenamefont {Aspelmeyer}\ \emph {et~al.}(2014)\citenamefont
  {Aspelmeyer}, \citenamefont {Kippenberg},\ and\ \citenamefont
  {Marquardt}}]{aspelmeyer_cavity-opto_review}%
  \BibitemOpen
  \bibfield  {author} {\bibinfo {author} {\bibfnamefont {M.}~\bibnamefont
  {Aspelmeyer}}, \bibinfo {author} {\bibfnamefont {T.~J.}\ \bibnamefont
  {Kippenberg}},\ and\ \bibinfo {author} {\bibfnamefont {F.}~\bibnamefont
  {Marquardt}},\ }\href {https://doi.org/10.1103/RevModPhys.86.1391} {\bibfield
   {journal} {\bibinfo  {journal} {Reviews of Modern Physics}\ }\textbf
  {\bibinfo {volume} {86}},\ \bibinfo {pages} {1391} (\bibinfo {year}
  {2014})}\BibitemShut {NoStop}%
\end{thebibliography}%

\end{document}


\title{A proposal for detecting the spin of a single electron in superfluid helium}

\author{Jinyong Ma}
\affiliation{Department of Physics, Yale University, New Haven, Connecticut 06520, USA}
\author{Yogesh S. S. Patil}
\affiliation{Department of Physics, Yale University, New Haven, Connecticut 06520, USA}
\author{Jiaxin Yu}
\affiliation{Department of Applied Physics, Yale University, New Haven, Connecticut 06520, USA}
\author{Yiqi Wang}
\affiliation{Department of Applied Physics, Yale University, New Haven, Connecticut 06520, USA}
\author{Jack G. E. Harris$^{}$}\email{jack.harris@yale.edu}
\affiliation{Department of Physics, Yale University, New Haven, Connecticut 06520, USA}
\affiliation{Department of Applied Physics, Yale University, New Haven, Connecticut 06520, USA}
\affiliation{Yale Quantum Institute, Yale University, New Haven, Connecticut 06520, USA}

\maketitle

This supplemental material presents detailed calculations for the analyses in the main paper. Section~\ref{sec:trap} provides the theory for acoustic trapping of an electron bubble. Section~\ref{sec:inter} discusses the optomechanical interaction between an electron bubble and an optical mode. Section~\ref{sec:damp} estimates the mechanical damping of the bubble motion. Section~\ref{sec:spin_det} gives details of the three protocols for single-spin detection proposed in the main paper. Section~\ref{sec:noise} analyzes the noise sources influencing the system performance. Section~\ref{sec:para} lists the various parameters used in the calculations.

\tableofcontents
\clearpage
  
\section{Sound wave trapping} \label{sec:trap}

\subsection{Energy and Young's modulus of electron bubbles}
We consider the following simple model for describing the total energy of an electron bubble (EB) in liquid helium~\cite{grimesInfrared1990, classenElectrons1998, konstantinovDetection2003,xingElectrons2020}
\begin{eqnarray} 
E_{{\rm tol}} & = & \frac{h^{2}}{8m_{\rm e}R^{2}}+4\pi R^{2}\sigma+\frac{4}{3}\pi R^{3}P \label{eq:el_E}
\end{eqnarray}
where the first term is the ground state energy of the confined electron, the second term is the surface tension energy of the bubble, and the last term represents the energy stored by the surrounding pressure (a more sophisticated model is described in Ref.~\cite{classenElectrons1998}). Here the electron mass is $m_{\rm e}$, the bubble radius is $R$, the surface tension of liquid helium is $\sigma = 0.375$ erg/cm$^2$~\cite{rocheMeasurement1997}, and the pressure experienced by the EB is $P$. When $P=0$, the EB radius is found by minimizing the total energy with respect to $R$, giving:
\begin{eqnarray}
R_0 & = & \left(\frac{h^{2}}{32\pi m_{\rm e}\sigma}\right)^{1/4} \approx 1.9 \ \rm{nm}
\end{eqnarray}

To find the response of the EB to an acoustic wave in the surrounding liquid, we rewrite Eq. (\ref{eq:el_E}) as

\begin{eqnarray}
P & = & \frac{h^{2}}{16\pi m_{\rm e}}\left(\frac{3}{4\pi}V_{\rm EB}\right)^{-5/3}-2\sigma\left(\frac{3}{4\pi}V_{\rm EB}\right)^{-1/3}
\end{eqnarray}

\noindent where $V_{\rm{EB}} = 4 \pi R^3 /3$ is the volume of the EB. It is then straightforward to calculate the  Young's modulus ($E_{{\rm EB}}$) of the EB: 

\begin{eqnarray}
E_{{\rm EB}} & = & -V_{\rm EB}\frac{\partial P}{\partial V_{\rm EB}} \label{eq:Eb} \approx 530  \ \rm{kPa}
\end{eqnarray}

\subsection{Potential of an acoustic trap} \label{sec:AcousticTrap}
An EB in a standing acoustic wave experiences a force exerted by the acoustic field. As described in the main text, our proposal employs this force to trap the EB. The potential energy of an EB in an acoustic wave (with wave vector $k$) can be written as~\cite{BruusAcoustofluidics2012a}

\begin{eqnarray}
U_{\rm ac} & = & V_{\rm EB}W_{{\rm ac}}\left[f_{1}\cos^{2}(kz)-\frac{3f_{2}}{2}\sin^{2}(kz)\right] \nonumber\\
 & = & V_{\rm EB}W_{{\rm ac}}\left[f_{1}-\Phi\sin^{2}(kz)\right] \label{eq:U_ac}
\end{eqnarray}

\noindent where $z$ is the coordinate of the EB along the cavity axis (see Fig. 1 of the main text), and

\begin{eqnarray}
f_{1} & = & 1-\frac{E_{{\rm He}}}{E_{{\rm EB}}}\\
f_{2} & = & \frac{2(\rho_{{\rm EB}}/\rho_{\rm He}-1)}{2\rho_{{\rm EB}}/\rho_{\rm He}+1}\\
\Phi & = & f_{1}+\frac{3}{2}f_{2}
\end{eqnarray}

\noindent Here $W_{{\rm ac}}$ is the acoustic energy density (see below), and the Young's modulus of the liquid He is $E_{{\rm He}} =\rho_{\rm He}c_{{\rm He}}^{2}$ where $c_{{\rm He}} = 238$ m/s is the velocity of sound in liquid He and $\rho_{\rm He} = 145$ kg/m$^3$ is its mass density. Since the interior of the EB is close to vacuum, we take its mass density $\rho_{{\rm EB}} = 0$. This gives $f_1 \approx -14$, $f_{2}\approx-2$, and $\Phi 
\approx -17$. The quantity $\Phi$ is known as the acoustophoretic contrast factor, and its sign determines whether the bubble is trapped at the pressure nodes or antinodes of the acoustic wave. Here $\Phi < 0$, so the EB will be trapped at pressure antinodes. 

The depth of the trapping potential in Eq.~(\ref{eq:U_ac}) is $U_0 = V_{\rm EB} \Phi W_{\rm ac}$. Throughout this paper, we assume $U_{0}/k_{{\rm B}}=300$ mK (chosen to be $10 \times$ greater than the assumed operating temperature $T = 30$ mK). This corresponds to an acoustic energy density of $W_{\rm ac} = 8.5$ \SI{}{J/m^3}. Generating such an acoustic wave is discussed in Section~\ref{sec:GenAc}.

To find the trap frequency of the EB, we expand $U_{\rm ac}$ to  second order in $z$ around its minimum at $z_0$, i.e. 
\begin{eqnarray}
U(z_{0}+\delta z) & \approx & U(z=z_{0})+\frac{1}{2}U^{\prime\prime}(z=z_{0})\delta z^{2}\nonumber\\
 & = & V_{\rm EB}W_{\rm ac}f_{1}+\frac{1}{2}k_{{\rm EB}}\delta z^{2} \label{eq:Uexp}
\end{eqnarray}
where $k_{{\rm EB}}=-2V_{\rm EB}W_{\rm ac}k^{2}\Phi\cos(2kz_{0})$ is the spring constant of the trap. This results in a trap frequency  $\omega_{{\rm EB}}=\sqrt{k_{{\rm EB}}/m_{{\rm EB}}} \approx 2 \pi \times 2.9$ MHz along the $z$-direction (here $m_{\rm EB} = 1.6\times10^{-24}$ kg is the bubble's effective mass~\cite{huangEffective2017}). 

\subsection{Generation of the acoustic trap with a piezoelectric actuator} \label{sec:GenAc}
The acoustic wave is generated by vibrating one of the cavity fibers with a piezoelectric actuator at frequency $\omega_{\rm ac}$. If the fiber displacement is $z_{{\rm f}}=z_{{\rm amp}}\sin(\omega_{{\rm ac}}t)$, the power in the sound wave ($\mathcal{P}_{\rm ac}$) emitted into the liquid helium is~\cite{HanSpecification2014a}

\begin{eqnarray}
\mathcal{P}_{{\rm ac}} & = & \sigma_{\rm rad}\rho_{\rm He}c_{{\rm He}}A\left\langle v^{2}\right\rangle \nonumber\\
 & = & \sigma_{\rm rad}\rho_{\rm He}c_{{\rm He}}A\left\langle \left(z_{{\rm amp}}\omega_{{\rm ac}}\cos(\omega_{{\rm ac}}t)\right)^{2}\right\rangle \nonumber\\
 & = & \frac{\sigma_{\rm rad}\rho_{\rm He}c_{{\rm He}}A\omega_{{\rm ac}}^{2}z_{{\rm amp}}^{2}}{2} \label{eq:P_ac-sigma}
\end{eqnarray}
where $\sigma_{\rm rad}$ is the radiation ratio (or efficiency), $v$ is the speed of the fiber tip of area $A$, and $\langle \cdot \rangle$ denotes the time average. $\mathcal{P}_{\rm ac}$ can also be written in terms of sound pressure or acoustic energy density

\begin{eqnarray}
\mathcal{P}_{{\rm ac}} & =2 A W_{{\rm ac}}c_{{\rm He}} \label{eq:P_ac-Wac}
\end{eqnarray}

We obtain the required vibration amplitude by combining Eqs.~\eqref{eq:P_ac-sigma}-\eqref{eq:P_ac-Wac}.
\begin{eqnarray}
z_{{\rm amp}} & = & \sqrt{\frac{4W_{{\rm ac}}}{\sigma_{\rm rad}\rho_{\rm He}\omega_{{\rm ac}}^{2}}}
\end{eqnarray}
Thus, to produce a sound wave that results in a trap depth corresponding to $300$ mK (see Sec I.B), the amplitude of the fiber motion would need to be $z_{\rm amp}\approx 240$ pm. However if the fiber motion is resonant with an acoustic mode of the cavity, this is reduced by a factor of the acoustic mode's quality factor $Q_{\rm{ac}}$. For the devices demonstrated in Ref.~\cite{shkarinQuantum2019a, KashkanovaSuperfluid2017a}, $Q_{\rm{ac}} \approx 10^5$. As a result, the required $z_{\rm amp}\approx 2.4$ fm. This amount of motion at comparable frequency is demonstrated, for example in \cite{leirsetHeterodyne2013}.

\subsection{Detection of the acoustic trap}
We first look at the acoustic energy stored in the fiber cavity $\mathcal{E}_{{\rm ac}}$, which is given by~\cite{shkarinQuantuma}
\begin{eqnarray}
\mathcal{E}_{{\rm ac}}  =  \frac{1}{2}E_{{\rm {\rm He}}}\int\left|\frac{\delta\rho_{\rm He}(r)}{\rho_{\rm He}}\right|^{2}dV = \frac{1}{2E_{{\rm He}}}\int\left|P(r)\right|^{2}dV \label{eq:Eac_drho}
\end{eqnarray}
where $P(r)$ is the pressure profile of the acoustic mode, and $\delta\rho_{\rm He}(r)$ is the corresponding density variation. Using $W_{{\rm ac}}=\frac{P_{0}^{2}}{4E_{{\rm He}}}$~\cite{BruusAcoustofluidics2012a} (where $P_0$ is the amplitude of the acoustic mode), and the well-known mode profiles for a Fabry-Perot cavity, we can write the time-averaged acoustic energy as

\begin{eqnarray}
\left\langle \mathcal{E}_{{\rm ac}}\right\rangle = \frac{1}{2}W_{{\rm ac}}V_{\rm ac} \label{eq:Eac_P0}
\end{eqnarray}
with $V_{{\rm ac}}$ denoting the mode volume of the acoustic cavity. 

The acoustic energy can also be written in terms of the phonon number $n_{\rm ac}$

\begin{eqnarray}
\left\langle \mathcal{E}_{{\rm ac}}\right\rangle  & = & \left( n_{\rm ac}+\frac{1}{2} \right) \hbar\omega_{{\rm ac}} \label{eq:Eac_n}
\end{eqnarray}

Equations~\eqref{eq:Eac_drho}-\eqref{eq:Eac_n} provide the phonon number $(n_{\rm ac}\thickapprox1.6\times10^{11})$ and the corresponding amplitude of the helium density variation $(\frac{\delta\rho_{\rm He}}{\rho_{{\rm He}}}\thickapprox2\times10^{-3})$ required to achieve the trap depth ($300$ mK) for the acoustic mode with $\omega_{\rm ac}/2\pi = 320 $ MHz, as considered in the main text.

The acoustic trap can be optically detected using the optomechanical
interaction between the optical and acoustic modes of
the fiber cavity. Under the drive of a monochromatic laser with frequency $\omega_{l}$ and amplitude $c_{l}$, the equations of motion for the amplitudes $c$ and $d$ of the optical and acoustic modes (with resonance frequencies $\omega_{\rm c}$ and $\omega_{\rm ac}$, and dampings $\kappa$ and $\gamma_{\rm ac}$) can be written in the rotating frame of $\omega_{l}$ as~\cite{kashkanovaOptomechanics, shkarinQuantuma},

\begin{eqnarray}
\dot{c} &=& -\left(\frac{\kappa}{2}-i\bar{\Delta}\right)c+\sqrt{\kappa_{\rm ex}}c_{l} , \\
\dot{d}	&=& -\left(\frac{\gamma_{{\rm ac}}}{2}+i\omega_{{\rm ac}}\right)d-ig_{0,{\rm ac}}c^{\dagger}c+d_{{\rm ext}}e^{-i\omega_{\rm ac}t} \label{eq:ac_d}
\end{eqnarray}
where 
$\Delta_l = \omega_l - \omega_{\rm c}$. Here we define the effective cavity detuning as $\bar{\Delta} = \Delta_{l}-ig_{0,{\rm ac}}(d+d^{\dagger})$ and denote $d_{\rm ext}$ as the external acoustic drive applied by the fiber vibration. The laser is injected into the cavity with input coupling rate $\kappa_{\rm ex}$, and the single-photon optomechanical coupling rate is $g_{\rm 0, ac}$. We consider a weak probe beam such that the radiation pressure on the acoustic wave is ignored. Eq.\ref{eq:ac_d} is then 

\begin{eqnarray}
\dot{d}&=&-\left(\frac{\gamma_{{\rm ac}}}{2}+i\omega_{{\rm ac}}\right)d+d_{{\rm ext}}e^{-i\omega_{\rm ac}t}
\end{eqnarray}
The solution to this equation is simply $d=d_{1}e^{-i\omega_{ac}t}$, where $d_{1}=2d_{{\rm ext}}/\gamma_{{\rm ac}}$. 

\begin{figure*}[htb] \centering
\includegraphics[width=0.7\textwidth]{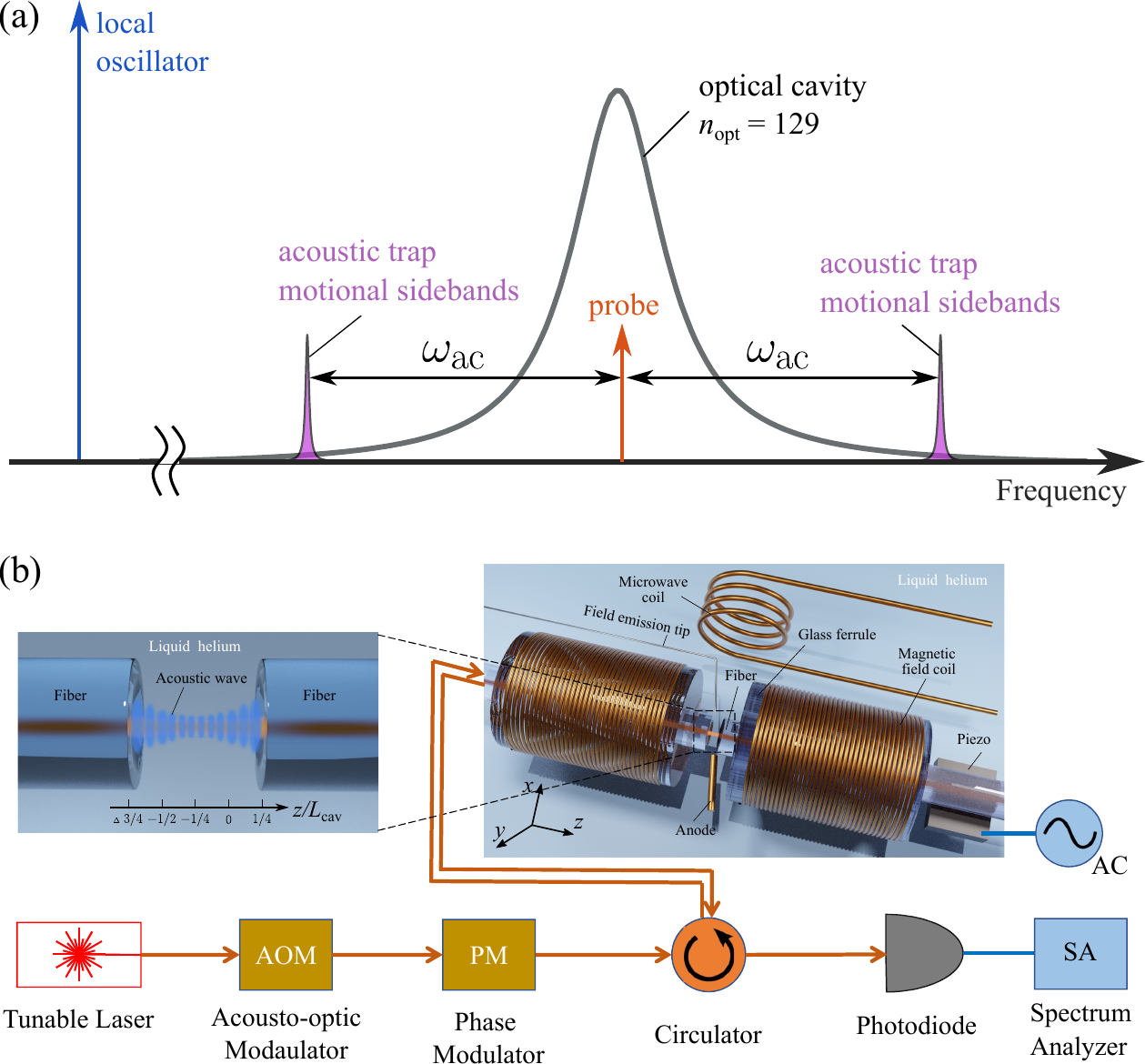} \caption{(a) Schematic of the laser frequencies for detecting the acoustic trap. The trap is formed by a density wave that oscillates with frequency $\omega_{\rm ac}$ and produces sidebands (at frequency $\omega_{\rm ac}$) on a ``probe'' laser tone that drives the optical cavity mode with $n_{\rm opt} =129$. These sidebands can be detected via standard heterodyne techniques using the ``local oscillator'' tone. (b) Schematic of the setup for detecting the acoustic trap. A laser beam passes through an acousto-optic modulator to produce the probe tone and the local oscillator. A phase modulator is used to add sidebands to the probe tone for the Pound-Drever-Hall (PDH) lock (not shown). Both tones are directed to the cavity via a circulator. One of the cavity fibers is driven via an ultrasonic actuator at frequency $\omega_{\rm ac}$ to produce the acoustic trap. The light reflected from the cavity is sent to a photodiode. Optical path: orange lines. Electrical path: blue lines.}
\label{fig:1}
\end{figure*}

In this analysis, the optical resonance frequency is modulated with a depth $g_{0,{\rm ac}}(d_1+d_1^{\dagger})$. With $d_1 = \sqrt{n_{\rm ac}}\approx 4 \times 10^5$ and the device parameters considered here ($\kappa / 2 \pi = 15$ MHz, $L_{\rm cav} = 100$ $\mu$m, $\mathcal{F}_{\rm opt} = 10^5$, $g_{\rm 0,ac} / 2 \pi = 3.6$ kHz), the cavity modulation depth $\approx 190 \kappa$. The approach for detecting the acoustic trap is presented in Fig.~\ref{fig:1}.

\section{Interaction between electron bubble and cavity optical field} \label{sec:inter}
Since the electron bubble radius ($R_0 \approx1.9$ nm) is much smaller than the optical wavelength ($\lambda_{\rm opt} = 1550$ nm), the Hamiltonian for a bubble coupling to the optical field (with frequency $\omega_{\rm c}$)
can be written as~\cite{KieselCavity2013a, NimmrichterMaster2010a}

\begin{eqnarray}
H  =  -\frac{1}{2}\alpha\left|E\right|^{2}  =  -\frac{\hbar\omega_{{\rm c}}\alpha}{2\varepsilon_{{\rm He}}V_{{\rm opt}}}f(z)a^{\dagger}a\\
\end{eqnarray}

\noindent where the EB polarizability is

\begin{eqnarray}
\alpha & = & 3\varepsilon_{{\rm He}}V_{{\rm EB}}\text{{\rm Re}}\left\{ \frac{n_{{\rm EB}}^{2}/n_{{\rm He}}^{2}-1}{n_{{\rm EB}}^{2}/n_{{\rm He}}^{2}+2}\right\} 
\end{eqnarray}
and $n_{\rm EB} = 1$ and $n_{\rm He} = 1.028$ are the refractive indices of the EB and the liquid He, $\epsilon_{\rm He} \approx n_{\rm He}^2$ is the permittivity of liquid He, and $V_{{\rm opt}}$ is the mode volume of the optical cavity. Here $f(z)$ is the profile of the optical cavity mode along the $z$ direction, given by
\begin{eqnarray}
f(z)  & = & \frac{1}{1+(z_0+z)^{2}/z_{{\rm R}}^{2}}\cos^{2}\left(k_{\rm{opt}}(z_0+z)\right)\nonumber\\
 & \approx & \cos^{2}\left(k_{\rm{opt}}z_{0}\right)-k_{\rm{opt}}\sin(2k_{\rm{opt}}z_{0})z-k_{\rm{opt}}^{2}\cos(2k_{\rm{opt}}z_{0})z^{2}+\mathcal{O}( z^{3})
\end{eqnarray}
with $ k_{\rm{opt}} = 2 \pi/\lambda_{\rm opt}$. The pre-factor $(1+(z/z_{\rm R})^2)^{-1}\approx 1$ for the cavities considered in Ref.~\cite{kashkanovaOptomechanics}. The term proportional to $z$ leads to the usual linear optomechanical interaction

\begin{eqnarray}
H & = & \hbar g_{0}a^{\dagger}a(b^{\dagger}+b)
\end{eqnarray}
with
\begin{eqnarray}
g_{0} & = & \frac{\omega_{{\rm c}}\alpha k_{\rm{opt}}}{2\varepsilon_{{\rm He}}V_{{\rm opt}}}\sin(2k_{\rm{opt}}z_{0})z_{{\rm zpf}} \label{eq:g0}
\end{eqnarray}
where $z_{{\rm zpf}}=\sqrt{\hbar/2m_{\rm EB}\omega_{{\rm EB}}}$ is the zero point
motion of the EB, $z = z_{\rm zpf} (b+b^\dagger)$, and $b$ denotes the annihilation operator of the EB motion. As discussed in the main paper, the equilibrium position of the trapped bubble corresponds to zero optomechanical coupling of the optical mode used to detect the acoustic trap. To maximize the coupling rate to the EB motion, one can use a different optical mode. The phase difference in the expression for $g_0$ (see Eq.~\ref{eq:g0}) for two optical modes of wavenumbers $k_{q+n}$ and $k_q$ (i.e. of longitudinal indices $q+n$ and $q$) at the same location $z_0$ is 

\begin{eqnarray}
2(k_{q+n}-k_{q})z_0 & = & \frac{2\pi n}{L_{\rm cav}}z_0
\end{eqnarray}

\noindent Considering an adjacent optical mode, i.e., $n=1$, the phase difference is $\pi/2$ at $z_0=L_{\rm cav}/4$, which maximizes the coupling rate.

\section{Damping Mechanism} \label{sec:damp}

An EB's motion may damped by a variety of mechanisms. We begin by noting some of the damping sources that are not expected to be relevant in the regimes considered for this study. Then we discuss the two sources that are expected to be relevant.

When the EB moves at speed (relative to the superfluid) greater than the critical velocity $v_L \approx 50$ m/s  it generates vortex rings and ripplons, leading to strong damping \cite{Allum1977,Nancolas1985}. This damping mechanism is not expected to be relevant, as even for the largest amplitude of motion considered in the main text ($z_{\rm{EB}}=100$ nm) the EB's speed $< 2$ m/s.

An EB may also be damped via interactions with remanent vortex lines. To avoid this effect, the number of vortex lines in the cavity volume will be minimized by filling the experimental cell via a Rollin film.

The EB may be damped by the thermal excitations in the superfluid. However, all thermal excitations are frozen out at the temperatures considered here ($T<100$ mK) except for phonons.

Lastly, $^3$He impurities are inevitably present at some concentration ($\approx 1$ ppm for natural He, and $\approx 10^{-6}$ ppm in isotopically purified He). In the regimes that are relevant for this work, the $^3$He atoms may be regarded as a gas whose collisions with an EB represent a source of damping.

In summary, the EB's damping is expected to be dominated by two sources: collisions with thermal phonons, and collisions with $^3$He impurities. We estimate the damping rates arising from each of these sources by using the results of Ref.~\cite{BaymMobility1969} (for the phonons) and Ref.~\cite{schwarzHe1972} (for the $^3$He).

In particular, the EB's phonon-limited mobility $\mu_{\rm p}$ is given by  Eq. (8) and Eq. (9) of Ref.~\cite{BaymMobility1969}, and the $^3$He-limited mobility $\mu_{\rm h}$ is given by the expressions in the first portion of Section III of Ref.~\cite{schwarzHe1972}. These mobilities are then converted to damping rates via $\gamma_{\rm p,h} = e/m\mu_{\rm p,h}$.

\begin{figure*}[htb] \centering
\includegraphics[width=0.6\textwidth]{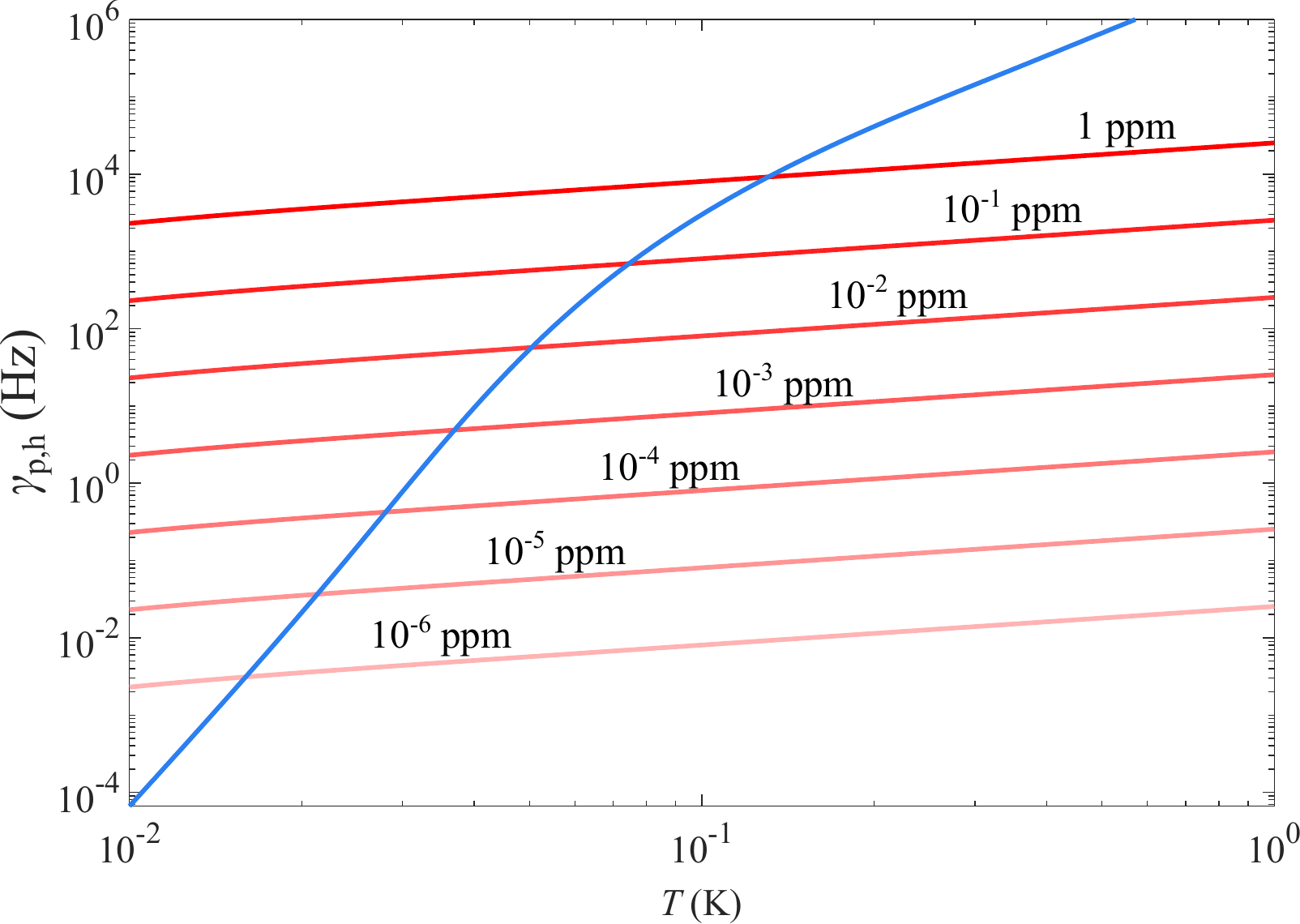} \caption{Damping rate of the electron bubble's motion.  The contribution from thermal phonons (blue) and various concentrations of $^3$He impurities (red) are shown, both as a function of temperature.}
\label{fig:2}
\end{figure*}

In Fig.~\ref{fig:2}, the blue curve shows the phonon contribution to the damping (calculated with $R_0 = 1.9$ nm, rather than the slightly smaller value assumed in Ref.~\cite{BaymMobility1969}), and the red curves show the $^3$He contribution for a range of $x_3$ (the concentration of $^3$He). The total EB damping $\gamma_{\rm EB} = \gamma_{\rm h} + \gamma_{\rm p}$ is converted to  $Q_{\rm EB} = \omega_{\rm EB} / \gamma_{\rm EB}$ and is shown in Fig. 3 of the main paper (as red curves) for the same values of $x_3$.

\section{Single spin detection} \label{sec:spin_det}

As discussed in the main text, the EB experiences a force along the $z$ direction given by:

\begin{eqnarray}
F_{{\rm mag}}  =\mu_{\rm S}G
\end{eqnarray}
where $\mu_{\rm S} = -g\mu_{\rm B}/2$ is the $z$-component of the electron's spin magnetic moment, $\mu_{\rm B}$ and $g$ denote the Bohr magneton and gyromagnetic ratio of the electron, and $G$ is the gradient (with respect to $z$) of the $z$ component of the magnetic field. 

In the following subsections, we will consider three approaches to obtain an alternating magnetic force that drives the bubble motion on resonance.

\subsection{Modulation of magnetic gradient}

Modulating the magnetic gradient at the trap frequency is the most straightforward approach to drive the EB's motion resonantly. In this case, the magnetic force is simply

\begin{eqnarray}
F_{{\rm mag}} & = & \mu_{\rm S}G \cos(\omega_{\rm EB} t+\phi) \label{eq:Mod_GB}
\end{eqnarray}

\noindent as described in the main text.

\subsection{Rabi oscillation}
In this subsection, we consider using Rabi oscillations to produce the oscillating force. We consider the following magnetic field:

\begin{eqnarray}
\vec{B} & = &  B_{0}\hat{z}+B_{1}\left(\cos(\omega t)\hat{x}-\sin(\omega t)\hat{y}\right)
\end{eqnarray}
where the static field $ B_{0}$ is used to induce the Zeeman splitting and $B_{1}$ is the amplitude of the microwaves that drive the spin resonance. The Hamiltonian of the system can be expressed as

\begin{eqnarray}
H  =  -\vec{\mu}\cdot\vec{B} = -\frac{\hbar}{2}\omega_{0}\sigma_{z}-\frac{\hbar}{2}\omega_{1}\left(\sigma_{x}\cos(\omega t)-\sigma_{y}\sin(\omega t)\right)
\end{eqnarray}
with $\omega_{0}  =  \frac{\mu_{\rm B} g}{\hbar} B_{0}$ and 
$\omega_{1}  =  \frac{\mu_{\rm B} g}{\hbar}B_{1}$. Here $\vec{\mu} = -\frac{1}{2}g\mu_{\rm B}(\sigma_x\hat{x}+\sigma_y\hat{y}+\sigma_z\hat{z})$ where $\sigma_{x,y,z}$ are the Pauli matrices. 

If the electron is initially in state $\left|\downarrow\right\rangle $, then the probability of the electron being found in spin $\left|\uparrow\right\rangle $ is given by

\begin{eqnarray}
P(t) & = & \left(\frac{\omega_{1}}{\Omega}\right)^{2}\sin^{2}\left(\frac{\Omega t}{2}\right)
\end{eqnarray}
Here $\Omega=\sqrt{\Delta^{2}+\omega_{1}^{2}}$ is the Rabi frequency where $\Delta=\omega-\omega_{0}$ is the detuning of the microwaves from the spin resonance. Assuming the initial magnetic moment is $\mu_{\rm S,0}$, we have the equation for magnetic force as follows

\begin{eqnarray}
F_{{\rm mag}} & = & \mu_{\rm S,0}P(t) G
\end{eqnarray}

\noindent It is straightforward to tune $B_1$ and $\Delta$ so that $\Omega = \omega_{\rm EB}$, producing an alternating force whose frequency is $\omega_{\rm EB}$.

\subsection{Adiabatic spin flip}
The third approach follows the protocol of Ref.~\cite{MaminDetection2003,RugarSingle2004}, and involves the modulation of the uniform field (at frequency $\omega_{\rm EB}$ and amplitude $B_{{\rm mod}}$) about a mean value $\bar B_0$. The frequency of the microwave field is $\omega_0 = \frac{\mu_{\rm B} g }{\hbar}\bar B_0$, so that it is resonant with the Zeeman splitting when $|\vec{B}|=\bar B_0$. Thus, the overall magnetic field is

\begin{eqnarray}
\vec{B} & = & \left(\bar B_{0}+B_{{\rm mod}}\sin(\omega_{\rm EB}t)\right)\hat{z}+B_{1}\left(\cos(\omega_0 t)\hat{x}-\sin(\omega_0 t)\hat{y}\right)
\end{eqnarray}

\noindent This results in the following Hamiltonian:

\begin{eqnarray}
H & = & -\vec{\mu}\cdot\vec{B}\nonumber =  -\frac{\hbar}{2}\left(\omega_{0}+\omega_{{\rm mod}}\sin(\omega_{\rm EB}t)\right)\sigma_{z}-\frac{\hbar}{2}\omega_{1}\left(\sigma_{x}\cos(\omega_0 t)-\sigma_{y}\sin(\omega_0 t)\right)\nonumber
\end{eqnarray}
 with $\omega_{{\rm mod}}=\frac{\mu_{\rm B}g}{\hbar}B_{{\rm mod}}$. For convenience, we rewrite this equation in the frame rotating at $\omega_0$:
\begin{eqnarray}
H^\prime & = & -\frac{\hbar}{2}\omega_{{\rm mod}}\sin(\omega_{\rm EB}t)\sigma_{z}-\frac{\hbar}{2}\omega_{1}\sigma_{x}  \label{eq:effH}
\end{eqnarray}

If the time-dependence of the Hamiltonian in Eq.~(\ref{eq:effH}) is slow enough that the spin state evolves adiabatically, then the z component of the magnetization is~\cite{MaminDetection2003a}

\begin{eqnarray}
\mu_{\rm s}(t) & = & \pm\frac{B_{{\rm mod}}\sin(\omega_{\rm EB}t)}{\sqrt{B_{1}^{2}+B_{{\rm mod}}^{2}\sin(\omega_{\rm EB}t)^{2}}}\mu_{{\rm eff}} \label{eq:mu_s}
\end{eqnarray}
where $\mu_{{\rm eff}}=\frac{1}{2}g\mu_{{\rm B}}$, and the magnetic force along the
$z$ direction is simply 
\begin{eqnarray}
F_{{\rm mag}} & = & \mu_{\rm s}(t)G
\end{eqnarray}\textbf{}

To estimate the probability that the spin evolves adiabatically under the  in Eq.~(\ref{eq:effH}), we model each passage through the avoided crossing (which occurs when $t = n \pi / \omega_{\rm EB}$, with $n$ an integer) using the Landau-Zener formula. Specifically, we replace $\sin(\omega_{\rm EB}t)$ in Eq.~(\ref{eq:effH}) with $\omega_{\rm EB}t$, so that the resulting Hamiltonian is equivalent to

\begin{eqnarray}
H^\prime & = & \frac{1}{2}\alpha t\sigma_{z}+{\rm Re}(H_{12})\sigma_{x}-{\rm Im}(H_{12})\sigma_{y}
\end{eqnarray}
where $H_{12}=-\frac{\hbar}{2}\omega_{1}$ and $\alpha=-\hbar\omega^{\prime}\omega_{{\rm EB}}$. 
In this case, the probability to find the system in the state $\left|\uparrow\right\rangle $  at $t=\infty$ given that the system is in the state $\left|\uparrow\right\rangle $  at $t=-\infty$ is the well-known expression

\begin{eqnarray}
P_{{\rm LZ}}(\uparrow\mid\uparrow)  = \exp\left(-2\pi\frac{\left|H_{12}\right|^{2}}{\hbar\left|\alpha\right|}\right)=\exp\left(-\frac{\pi}{2}\frac{\omega_{1}^{2}}{\omega_{\rm EB}\omega_{{\rm mod}}}\right) \label{eq:P_LZ}
\end{eqnarray}

As a result, the probability that the spin state evolves adiabatically (and hence the Eq.~(\ref{eq:mu_s}) holds) is $1-P_{{\rm LZ}}(\uparrow\mid\uparrow)$. This probability approaches unity when 
$\omega_{\rm EB}\omega_{{\rm mod}}  \ll  \omega_{1}^{2}$.

Equation~(\ref{eq:P_LZ}) is used to calculate the probability of adiabatic evolution given in Fig. 4(c,d) of the main text.

\section{Noise analysis} \label{sec:noise}
In this section, we start with the equations of motion for the EB and the optical mode and analyse the noise sources associated with measuring the EB displacement.
\subsection{Optomechanical interaction}
The Hamiltonian describing the system can be written as
\begin{eqnarray}
H & = & \hbar\omega_{c}a^{\dagger}a+\hbar\omega_{\rm EB}b^{\dagger}b+\hbar g_{\rm 0}a^{\dagger}a(b^{\dagger}+b)+F_{{\rm mag}}x_{{\rm zpf}}(b^{\dagger}+b)
\end{eqnarray}
The first three terms characterize the optical mode of the fiber cavity, the EB motion, and their interaction, respectively. The last term describes the magnetic force imprinted from the electron spin~\cite{ma_OMIT-mechanical-drive_opto}. Adding the optical ($\kappa$) and mechanical ($\gamma_{\rm EB}$) damping terms and environmental fluctuations, the Heisenberg equations of motion are

\begin{eqnarray}
\dot{a} & = & -\left(\frac{\kappa}{2}+i\omega_{c}\right)a-ig_{\rm 0}(b^{\dagger}+b)a+\sqrt{\kappa_{\rm in}}\xi_{\rm in}+\sqrt{\kappa_{\rm ex}}(a_{\rm ex}+\xi_{\rm ex}) \label{eq:a}\\
\dot{b} & = & -\left(\frac{\gamma_{\rm EB}}{2}+i\omega_{\rm EB}\right)b-ig_{\rm 0}a^{\dagger}a-\frac{i}{\hbar}F_{{\rm mag}}x_{{\rm zpf}}+\sqrt{\gamma_{\rm EB}}\eta \label{eq:b}
\end{eqnarray}
where $\xi_{\rm in}$ and $\xi_{\rm ex}$ denote the optical noises coupled to the system via internal and external channels with the corresponding coupling rates $\kappa_{\rm in}$ and $\kappa_{\rm ex}$. We have also included the mechanical noise $\eta$ that comes from the thermal bath.  The noise operators' correlations are

\begin{eqnarray} \label{eq:noise_t}
\left\langle \xi_{i}(t)\xi_{j}(t^{'})\right\rangle  & = & 0\\
\left\langle \xi_{i}^{\dagger}(t)\xi_{j}(t^{'})\right\rangle  & = & 0\\
\left\langle \xi_{i}(t)\xi_{j}^{\dagger}(t^{'})\right\rangle  & = & \delta_{ij}\delta(t-t^{'})\\
\left\langle \eta(t)\eta(t^{'})\right\rangle  & = & 0\\
\left\langle \eta^{\dagger}(t)\eta(t^{'})\right\rangle  & = & n_{\rm th}\delta(t-t^{'})\\
\left\langle \eta(t)\eta^{\dagger}(t^{'})\right\rangle  & = & (n_{\rm th}+1)\delta(t-t^{'})
\end{eqnarray} 

where $n_{\text{th}}$ is the thermal phonon occupancy of the EB's motion.

We consider a monochromatic optical drive, i.e., $a_{\rm ex}=a_{l}e^{-i\omega_{l}t}$. In the frame rotating at $\omega_{c}$, we rewrite Eqs.~\eqref{eq:a} and~\eqref{eq:b} as

\begin{eqnarray}
\dot{a} & = & -\frac{\kappa}{2}a-ig_{\rm 0}(b^{\dagger}+b)a+\sqrt{\kappa_{\rm in}}\xi_{\rm in}+\sqrt{\kappa_{\rm ex}}\xi_{\rm ex}+\sqrt{\kappa_{\rm ex}}a_{l}e^{-i\Delta_{l}t}\\
\dot{b} & = & -\left(\frac{\gamma_{\rm EB}}{2}+i\omega_{\rm EB}\right)b-i g_{\rm 0} a^{\dagger}a-\frac{i}{\hbar}F_{{\rm mag}}x_{{\rm zpf}}+\sqrt{\gamma_{\rm EB}}\eta
\end{eqnarray}
where $\Delta_{l}=\omega_{l}-\omega_{c}$. Applying the linearization:
$a\rightarrow a_{0}+a$ and $b\rightarrow b_{0}+b$, we get the zeroth-order
equations
 
\begin{eqnarray}
\dot{a_{0}} & = & -\frac{\kappa}{2}a_{0}-ig_{\rm 0}(b_{0}^{*}+b_{0})a_{0}+\sqrt{\kappa_{\rm ex}}a_{l}e^{-i\Delta_{l}t}\\
\dot{b_{0}} & = & -\left(\frac{\gamma_{\rm EB}}{2}+i\omega_{\rm EB}\right)b_{0}-iga_{0}^{*}a_{0}
\end{eqnarray}
and the first-order equations
\begin{eqnarray}
\dot{a} & = & -\frac{\kappa}{2}a-ig_{\rm 0}(a_{0}b^{\dagger}+a_{0}b+b_{0}^{*}a+b_{0}a)+\sqrt{\kappa_{\rm in}}\xi_{\rm in}+\sqrt{\kappa_{\rm ex}}\xi_{\rm ex}\\
\dot{b} & = & -\left(\frac{\gamma_{\rm EB}}{2}+i\omega_{\rm EB}\right)b-ig_{\rm 0}(a_{0}^{*}a+a^{\dagger}a_{0})-\frac{i}{\hbar}F_{{\rm mag}}x_{{\rm zpf}}+\sqrt{\gamma_{\rm EB}}\eta
\end{eqnarray}

The zero-order solutions for the optical and mechanical modes are:
\begin{eqnarray}
a_{0} & = & \frac{\sqrt{\kappa_{\rm ex}}a_{l}}{\kappa/2-i\bar{\Delta}}e^{-i\Delta_{l}t} \\
b_{0} & = & \frac{-iga_{0}^{*}a_{0}}{\frac{\gamma_{\rm EB}}{2}+i\omega_{\rm EB}} 
\end{eqnarray}
where $\bar{\Delta}=\Delta_{l}-g_{0}(b_{0}^{*}+b_{0})$. The reduced first-order
equations are

\begin{eqnarray}
\dot{a} & = & -\frac{\kappa}{2}a-ig_{\rm 0}a_{0}(b^{\dagger}+b)+\sqrt{\kappa}\xi \label{eq:a1}\\
\dot{b} & = & -\left(\frac{\gamma_{\rm EB}}{2}+i\omega_{\rm EB}\right)b-ig_{\rm 0}(a_{0}^{*}a+a^{\dagger}a_{0})-\frac{i}{\hbar}F_{{\rm mag}}x_{{\rm zpf}}+\sqrt{\gamma_{\rm EB}}\eta \label{eq:b1}
\end{eqnarray}
where we have introduced a combined vacuum noise operator,

\begin{eqnarray}
\xi & = & (\sqrt{\kappa_{\rm in}}\xi_{\rm in}+\sqrt{\kappa_{\rm ex}}\xi_{\rm ex})/\sqrt{\kappa}
\end{eqnarray}

Equations~\eqref{eq:a1} and \eqref{eq:b1} can be written in the frequency domain as follows
\begin{eqnarray}
a[\omega] & = & \chi_{c}[\omega]\left(-ig_{\rm 0}a_{0}(b^{\dagger}[\omega-\Delta_{l}]+b[\omega-\Delta_{l}])+\sqrt{\kappa}\xi[\omega]\right)\\
b[\omega] & = & \chi_{\rm m}[\omega]\left(-ig_{\rm 0}(a_{0}^{*}a[\omega+\Delta_{l}]+a_{0}a^{\dagger}[\omega-\Delta_{l}])-\frac{i}{\hbar}x_{{\rm zpf}}F_{{\rm mag}}[\omega]+\sqrt{\gamma_{\rm EB}}\eta[\omega]\right)
\end{eqnarray}
Here we use the noise operators' correlations in the frequency domain:
\begin{eqnarray}
\left\langle \xi(\omega)\xi(-\omega)\right\rangle  & = & 0\\
\left\langle \xi^{\dagger}(\omega)\xi(-\omega)\right\rangle  & = & 0\\
\left\langle \xi(\omega)\xi^{\dagger}(-\omega)\right\rangle  & = & 1\\
\left\langle \eta(\omega)\eta(-\omega)\right\rangle  & = & 0\\
\left\langle \eta^{\dagger}(\omega)\eta(-\omega)\right\rangle  & = & n_{th} \label{eq:eta_w1}\\
\left\langle \eta(\omega)\eta^{\dagger}(-\omega)\right\rangle  & = & n_{th}+1 \label{eq:eta_w2}
\end{eqnarray}
Note that $x^{\dagger}[\omega]=(x[-\omega])^{\dagger}$. We also define optical and mechanical susceptibility as
\begin{eqnarray}
\chi_{c}[\omega] & = & \left(\frac{\kappa}{2}-i\omega\right)^{-1}\\
\chi_{\rm m}[\omega] & = & \left(\frac{\gamma_{\rm EB}}{2}-i(\omega-\omega_{\rm EB})\right)^{-1}
\end{eqnarray}

The solution to \eqref{eq:b1} is easily obtained as
\begin{eqnarray}
b[\omega] & = & \chi_{{\rm eff}}[\omega]\left(-i\frac{x_{{\rm zpf}}}{\hbar}F_{{\rm RPSN}}[\omega]-i\frac{x_{{\rm zpf}}}{\hbar}F_{{\rm mag}}[\omega]+\sqrt{\gamma_{\rm EB}}\eta[\omega]\right) \label{eq:bw}\\
b^{\dagger}[\omega] & = & (\chi_{{\rm eff}}[-\omega])^{*}\left(i\frac{x_{{\rm zpf}}}{\hbar}F_{{\rm RPSN}}[\omega]+i\frac{x_{{\rm zpf}}}{\hbar}F_{{\rm mag}}^{\dagger}[\omega]+\sqrt{\gamma_{\rm EB}}\eta^{\dagger}[\omega]\right) \label{eq:b_dagger_w}
\end{eqnarray}
with
\begin{eqnarray}
\chi_{{\rm eff}}^{-1}[\omega] & = & \chi_{\rm m}^{-1}[\omega]+i\Sigma_{\rm opt}[\omega]\\
\Sigma_{\rm opt}[\omega] & = & ig_{0}^{2}\left(\left|a_{0}\right|^{2}\chi_{c}[\omega-\Delta_{l}]-\left|a_{0}\right|^{2}\chi_{c}[\omega+\Delta_{l}]\right)\\
F_{{\rm RPSN}}[\omega] & = & \frac{\hbar}{x_{{\rm zpf}}}g_{0}\sqrt{\kappa}\left(a_{0}^{*}\chi_{c}[\omega+\Delta_{l}]\xi[\omega+\Delta_{l}]+a_{0}\chi_{c}[\omega-\Delta_{l}]\xi^{\dagger}[\omega-\Delta_{l}])\right) \label{eq:F_RPSN}
\end{eqnarray}
where $\chi_{{\rm eff}}$ is the effective mechanical susceptibility modified by the optical backaction. The optical spring $\Omega_{{\rm opt}}$ and optical damping $\Gamma_{{\rm opt}}$ can be extracted from $\Sigma_{\rm opt}[\omega]$
\begin{eqnarray}
\Omega_{{\rm opt}} & = & {\rm Re}\left[\Sigma_{\rm opt}[\omega]\right]\\
\Gamma_{{\rm opt}} & = & -2{\rm Im}\left[\Sigma_{\rm opt}[\omega]\right]
\end{eqnarray}

The radiation pressure shot noise $F_{{\rm RPSN}}$ (see Eq.~\eqref{eq:F_RPSN}) originates from the backaction of the optical fluctuations. Its power spectral density (PSD) is found using the Wiener-Khinchin theorem

\begin{eqnarray}
S_{FF}^{{\rm RPSN}}[\omega] & = & \left\langle F_{{\rm RPSN}}[\omega]F_{{\rm RPSN}}[-\omega]\right\rangle \nonumber\\
 & = & \frac{\hbar^{2}g_{0}^{2}\left|a_{0}\right|^{2}}{x_{{\rm zpf}}^{2}}\frac{\kappa}{\kappa^{2}/4+(\omega+\Delta_{l})^{2}}
\end{eqnarray}

With the correlations of thermal noise operators given in Eqs.~\eqref{eq:eta_w1}-\eqref{eq:eta_w2}, we can easily get the PSD of thermal fluctuation

\begin{eqnarray}
S_{F^{\dagger}F}^{{\rm th}}[\omega] & = & \frac{\hbar^{2}}{x_{{\rm zpf}}^{2}}\left\langle \sqrt{\gamma_{\rm EB}}\eta^{\dagger}[\omega]\sqrt{\gamma_{\rm EB}}\eta[-\omega]\right\rangle 
  =  \frac{\hbar^{2}}{x_{{\rm zpf}}^{2}}\gamma_{\rm EB}n_{{\rm th}} \label{eq:S_FdF}\\
S_{FF^{\dagger}}^{{\rm th}}[\omega] & = & \frac{\hbar}{x_{{\rm zpf}}}\left\langle \sqrt{\gamma_{\rm EB}}\eta[\omega]\sqrt{\gamma_{\rm EB}}\eta^{\dagger}[-\omega]\right\rangle
  =  \frac{\hbar^{2}}{x_{{\rm zpf}}^{2}}\gamma_{\rm EB}(n_{{\rm th}}+1) \label{eq:S_FFd}
\end{eqnarray}

The difference in $S_{F^{\dagger}F}^{{\rm th}}[\omega]$ and $S_{FF^{\dagger}}^{{\rm th}}[\omega]$ will lead to the well-known quantum phenomenon called sideband asymmetry~\cite{aspelmeyer_cavity-opto_review}. These two equations can be rewritten as follows for large thermal occupations 

\begin{figure*}[htb] \centering
\includegraphics[width=0.7\textwidth]{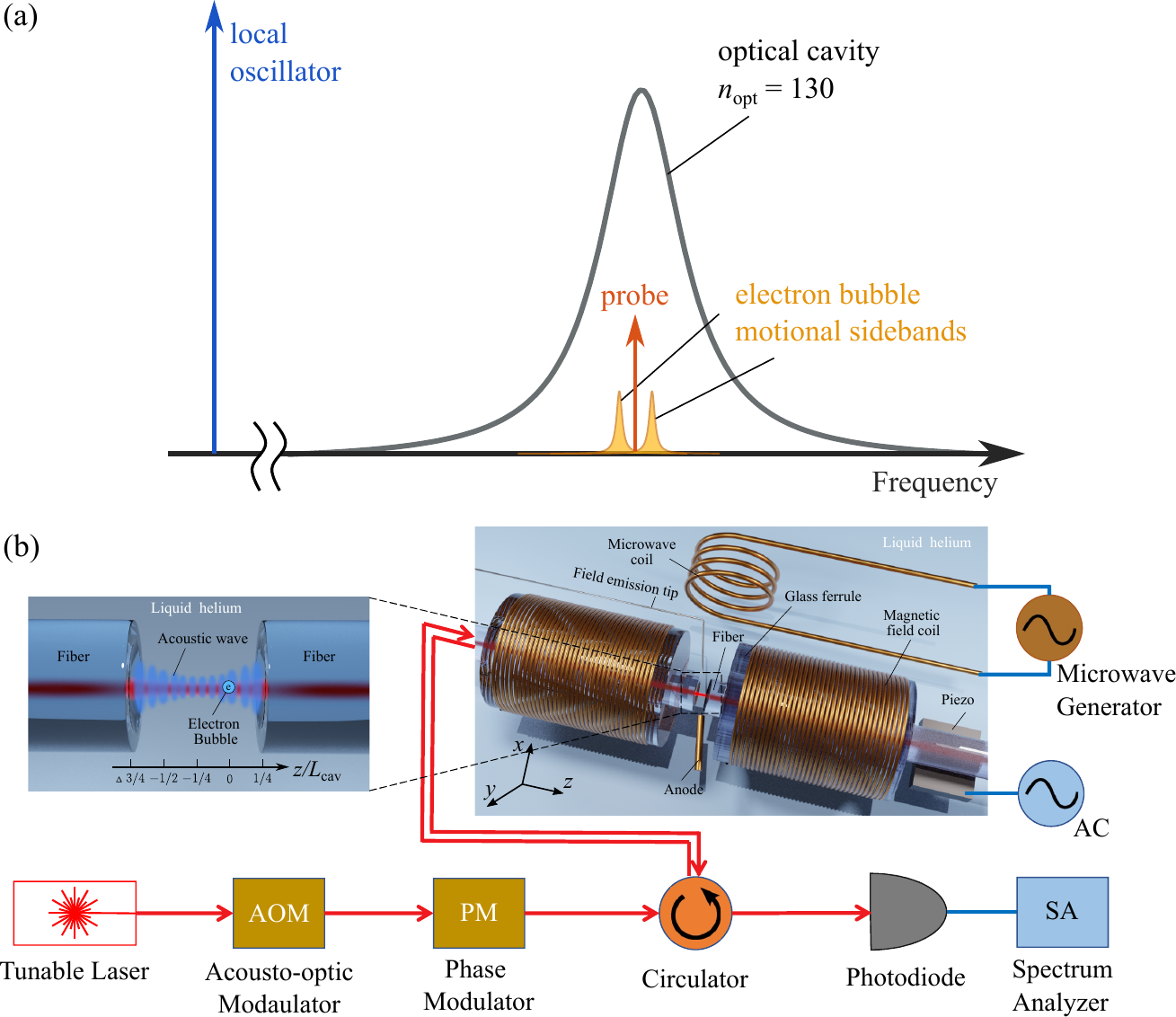} \caption{(a) Schematic of the laser frequencies for detecting the EB's motion. A ``probe'' laser tone  drives the optical cavity mode with $n_{\rm opt} =130$. Oscillations of the EB in the acoustic trap produce sidebands at frequencies $\pm \omega_{\rm{EB}}$. These sidebands are detected via standard heterodyne techniques using the ``local oscillator'' tone. (b) Schematic setup for electron spin detection. An EB produced from the field emission tip is trapped by the acoustic wave. The electron's spin is manipulated via the microwave coil, which is driven by a microwave generator. The spin state is transduced to the EB's displacement via the magnetic field gradient, which is generated by the field coil. A laser beam passes through an acousto-optic modulator to produce the probe tone and the local oscillator. The probe tone passes through a phase modulator to add sidebands for the PDH lock (not shown). Both tones are directed to the cavity via a circulator. The light reflected from the cavity is sent to a photodiode. Optical path: red lines. Electrical path: blue lines.} 
\label{fig:3}
\end{figure*}

\begin{eqnarray}
S_{F^{\dagger}F}^{{\rm th}}[\omega] \approx S_{FF^{\dagger}}^{{\rm th}}[\omega] \approx  2k_{B}Tm_{\rm EB}\gamma_{\rm EB}
\end{eqnarray}
Here we have used $n_{{\rm th}} \approx k_{B}T/\hbar\omega_{\rm EB} \gg 1$, as $T=30$ mK and $\omega_{\rm EB} = 2\pi \times 2.9$ MHz corresponds to $n_{{\rm th}} \approx 215$.
Equations~\eqref{eq:bw} and ~\eqref{eq:b_dagger_w} suggest the following PSD for the mechanical mode
\begin{eqnarray}
S_{b^{\dagger}b}[\omega] & = & \left|\chi_{{\rm eff}}[-\omega]\right|^{2}\frac{x_{{\rm zpf}}^{2}}{\hbar^{2}}\left(S_{FF}^{{\rm RPSN}}[\omega]+S_{FF}^{{\rm mag}}[\omega]+S_{F^{\dagger}F}^{{\rm th}}[\omega]\right)\\
S_{bb^{\dagger}}[\omega] & = & \left|\chi_{{\rm eff}}[\omega]\right|^{2}\frac{x_{{\rm zpf}}^{2}}{\hbar^{2}}\left(S_{FF}^{{\rm RPSN}}[\omega]+S_{FF}^{{\rm mag}}[\omega]+S_{FF^{\dagger}}^{{\rm th}}[\omega]\right)
\end{eqnarray}
Because $x[\omega]=x_{{\rm zpf}}(b[\omega]+b^{\dagger}[\omega])$, we obtain the PSD of the mechanical displacement
\begin{eqnarray}
S_{xx}[\omega] & = & x_{{\rm zpf}}^{2}\left(S_{bb^{\dagger}}[\omega]+S_{b^{\dagger}b}[\omega]\right)\nonumber\\
 & = & \left|\chi_{{\rm eff}}[-\omega]\right|^{2}\frac{1}{4m^{2}\omega_{\rm EB}^{2}}\left(S_{FF}^{{\rm RPSN}}[\omega]+S_{FF}^{{\rm mag}}[\omega]+S_{F^{\dagger}F}^{{\rm th}}[\omega]\right)\nonumber\\
 &  & +\left|\chi_{{\rm eff}}[\omega]\right|^{2}\frac{1}{4m^{2}\omega_{\rm EB}^{2}}\left(S_{FF}^{{\rm RPSN}}[\omega]+S_{FF}^{{\rm mag}}[\omega]+S_{FF^{\dagger}}^{{\rm th}}[\omega]\right)
\end{eqnarray}

Having found the solution for $b[\omega]$, the solution of the intracavity field in Eq. \eqref{eq:a1} is obtained.

With the input-output relation $a_{{\rm out}}[\omega]  =  \sqrt{\kappa_{{\rm ex}}}a[\omega]$, we find the PSD for the cavity output as 

\begin{eqnarray}
S_{a^{\dagger}a}^{{\rm out}}[\omega] & = & \kappa_{{\rm ex}}\left\langle a^{\dagger}[\omega]a[-\omega]\right\rangle \\
 & = & \kappa_{{\rm ex}}\left|\chi_{c}[-\omega]\right|^{2}g_{\rm 0}^{2}\left|a_0\right|^{2}\left(S_{b^{\dagger}b}[\omega+\Delta_{l}]+S_{bb^{\dagger}}[\Delta_{l}+\omega])\right)
\end{eqnarray}

\begin{figure*}[htb] \centering
\includegraphics[width=0.65\textwidth]{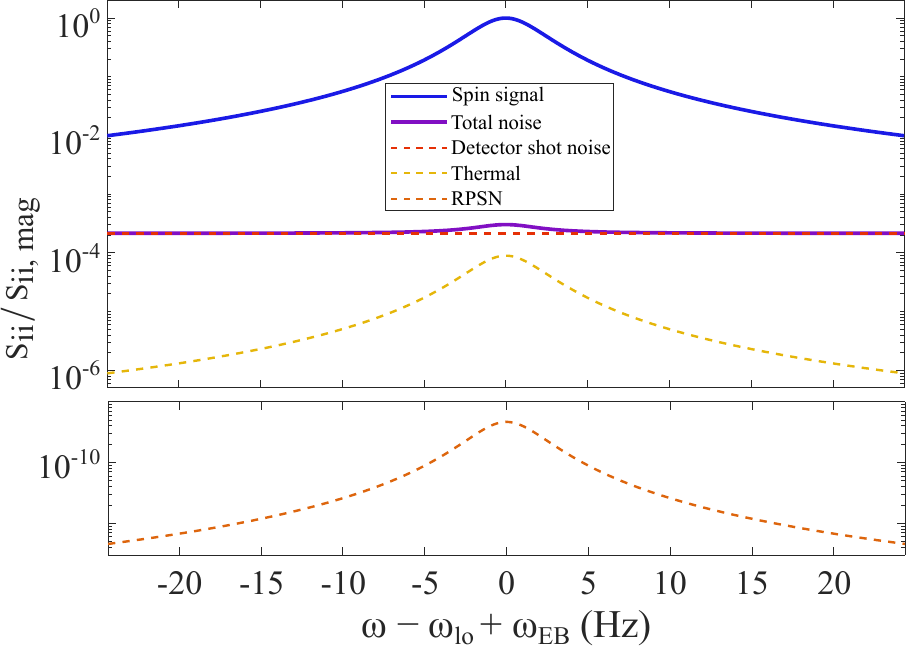} \caption{The power spectral density of the photodiode current $S_{ii}[\omega]$, normalized to $S_{ii,\rm{mag}}[0]$ (the magnetic force signal on resonance), assuming a measurement bandwidth $b = 1$ Hz.  The contributions to $S_{ii}$ from RPSN, the EB's thermal motion, and the detector shot noise are each shown. The parameters used here are: $Q_{\rm EB} = 6\times10^5$, $T = 30$~mK, $P=3$~\textmu W, $\mathcal{F} = 10^5$.} 
\label{fig:4}
\end{figure*}

\subsection{Heterodyne detection}
Here we consider optical heterodyne detection. The optical field $a_{{\rm det}}$ incident on a photodiode includes the cavity output and local oscillator, as shown in Fig.~\ref{fig:3}, which is given by

\begin{eqnarray}
a_{{\rm det}} & = & a_{{\rm lo}}e^{i\omega_{{\rm lo}}t}+a
\end{eqnarray}
where $a_{{\rm lo}}$ and $\omega_{{\rm lo}}$ denote the amplitude and frequency of the local oscillator. The autocorrelation of the photocurrent is

\begin{eqnarray}
G_{ii}(t,\tau) & = & \left\langle i(t+\tau/2)i(t-\tau/2)\right\rangle \nonumber\\
 & = & G^{2}\left\langle :a_{{\rm det}}^{\dagger}(t+\tau/2)a_{{\rm det}}(t+\tau/2)a_{{\rm det}}^{\dagger}(t-\tau/2)a_{{\rm det}}(t-\tau/2):\right\rangle +G^{2}\left\langle a_{{\rm det}}^{\dagger}(t)a_{{\rm det}}(t)\right\rangle \delta(\tau) \nonumber\\
 & \approx & G^{2}\left|a_{{\rm lo}}\right|^{4}+G^{2}\left|a_{{\rm lo}}\right|^{2}\left\langle a^{\dagger}(t+\tau/2)a(t+\tau/2)\right\rangle +G^{2}\left|a_{{\rm lo}}\right|^{2}\left\langle a^{\dagger}(t-\tau/2)a(t-\tau/2)\right\rangle \nonumber\\
 &  & +G^{2}\left|a_{{\rm lo}}\right|^{2}e^{i\omega_{{\rm lo}}\tau}\left\langle a^{\dagger}(t+\tau/2)a(t-\tau/2)\right\rangle +G^{2}\left|a_{{\rm lo}}\right|^{2}e^{-i\omega_{{\rm lo}}\tau}\left\langle a^{\dagger}(t-\tau/2)a(t+\tau/2)\right\rangle \nonumber\\
 &  & +G^{2}a_{{\rm lo}}^{2}e^{2i\omega_{{\rm lo}}t}\left\langle a^{\dagger}(t+\tau/2)a^{\dagger}(t-\tau/2)\right\rangle +G^{2}\left(a_{{\rm lo}}^{*}\right)^{2}e^{-2i\omega_{{\rm lo}}t}\left\langle a(t+\tau/2)a(t-\tau/2)\right\rangle \nonumber\\
 &  & +G^{2}\left|a_{{\rm lo}}\right|^{2}\delta(\tau)
\end{eqnarray}
where $G$ is the gain of the detector and the last term represents
the detector shot noise. The PSD of the photocurrent is the Fourier
transform of the correlation

\begin{eqnarray}
S_{ii}[\omega] & = & G^{2}\left|a_{{\rm lo}}\right|^{2}\left(S_{a^{\dagger}a}[\omega_{{\rm lo}}+\omega]+S_{a^{\dagger}a}[\omega_{{\rm lo}}-\omega]+1\right)\nonumber\\
 & = &G^{2}\left|a_{{\rm lo}}\right|^{2}\bigg(g^{2}\left|a_0\right|^{2}\left|\chi_{c}[-(\omega_{{\rm lo}}+\omega)]\right|^{2}\big(S_{b^{\dagger}b}[\omega_{{\rm lo}}+\omega+\Delta_{l}]+S_{bb^{\dagger}}[\Delta_{l}+\omega_{{\rm lo}}+\omega]\big)\nonumber\\
 &  & +g^{2}\left|a_0\right|^{2}\left|\chi_{c}[-(\omega_{{\rm lo}}-\omega)]\right|^{2}\big(S_{b^{\dagger}b}[\omega_{{\rm lo}}-\omega+\Delta_{l}]+S_{bb^{\dagger}}[\Delta_{l}+\omega_{{\rm lo}}-\omega]\big)+1\bigg)
\end{eqnarray}

Figure.~\ref{fig:4} shows the contribution to $S_{ii}$ from each noise source, as well as their total.

\section{Table of parameters} \label{sec:para}
{
\centering
\begin{tabular}{|c|c|}
\hline 
\multicolumn{2}{|c|}{\textbf{Optical cavity} }\tabularnewline
\hline 
Optical resonant wavelength ($\lambda_{{\rm opt}}$) & 1550 nm\tabularnewline
\hline 
Optical cavity linewidth ($\kappa/2\pi$) & 15 MHz\tabularnewline
\hline 
Optical cavity input coupling rate ($\kappa_{\rm ex}$) & $0.44 \kappa$\tabularnewline
\hline 
Cavity Finesse ($\mathcal{F}$) & $10^{5}$\tabularnewline
\hline 
Cavity length ($L_{{\rm cav}}$) & 100 $\mu$m\tabularnewline
\hline 
Cavity waist ($w_{{\rm cav}}$) & 5 $\mu$m\tabularnewline
\hline 
Operating temperature ($T$) & 30 mK\tabularnewline
\hline 
\multicolumn{2}{|c|}{}\tabularnewline
\hline 
\multicolumn{2}{|c|}{\textbf{Acoustic trap}}\tabularnewline
\hline 
Acoustic wavelength ($\lambda_{{\rm ac}}$) & 775 nm\tabularnewline
\hline 
Acoustic resonant frequency ($\omega_{{\rm ac}}/2\pi$) & 320 MHz\tabularnewline
\hline 
Acoustic optomechanical coupling ($g_{0,{\rm ac}}/2\pi$) & 3.6 kHz\tabularnewline
\hline 
Quality factor of acoustic mode ($Q_{{\rm ac}}$) & $10^{5}$\tabularnewline
\hline 
Acoustic trap depth ($U_{0}/k_{{\rm B}}$)  & 300 mK\tabularnewline
\hline 
Fiber vibration amplitude ($z_{{\rm amp}}$) & 2.4 fm\tabularnewline
\hline 
Acoustic energy density ($W_{{\rm ac}}$) & 8.5 J/m$^3$\tabularnewline
\hline 
Phonon number ($n_{{\rm ac}}$) & $1.6\times10^{11}$\tabularnewline
\hline 
Helium density variation ($\delta\rho_{{\rm He}}/\rho_{{\rm He}}$) & $2\times10^{-3}$\tabularnewline
\hline 
Optical mode for acoustic wave detection ($n_{{\rm opt}}$) & 129\tabularnewline
\hline 
Acoustic mode number ($n_{{\rm ac}}$) & 258\tabularnewline
\hline 
\multicolumn{2}{|c|}{}\tabularnewline
\hline 
\multicolumn{2}{|c|}{\textbf{Electron bubble}}\tabularnewline
\hline 
Electron bubble mass ($m_{{\rm EB}}$) & $1.6\times10^{-24}$ kg\tabularnewline
\hline 
Electron bubble radius ($R_{0}$) & 1.9 nm\tabularnewline
\hline 
Trap frequency ($\omega_{{\rm EB}}/2\pi$) & 2.9 MHz\tabularnewline
\hline 
Young's modulus of the EB ($E_{\rm EB}$) & 530 kPa\tabularnewline
\hline 
Optomechanical coupling rate ($g_{0}/2\pi$) & 0.1 Hz\tabularnewline
\hline 
Cavity mode for EB detection ($n_{{\rm opt}}$) & 130\tabularnewline
\hline 
\end{tabular} \par
}

\vspace{10mm}

\bibliographystyle{apsrev4-2}
\bibliography{supp}